\documentclass[10pt,journal]{IEEEtran}

\usepackage{amsmath,amssymb,mathtools}
\usepackage{graphicx}
\usepackage{booktabs}
\usepackage{multirow}
\usepackage{array}
\usepackage{tabularx}
\usepackage{makecell}
\usepackage{xcolor}
\usepackage{cite}
\usepackage{url}
\usepackage{enumitem}
\usepackage{pifont}
\usepackage{placeins}
\usepackage{tikz}
\usetikzlibrary{
    arrows.meta,
    positioning,
    fit,
    calc,
    shapes.geometric,
    decorations.pathreplacing
}
\newcommand{\cmark}{\ding{51}}
\newcommand{\xmark}{\ding{55}}

\newcommand{\KDF}{\operatorname{KDF}}

\title{Cross-Layer Roots of Trust: Integrating Biometrics, PUFs, and Hardware Obfuscation}

\author{Nima Karimian}

\begin{document}
\maketitle

\begin{abstract}
Modern cyber--physical, Internet-of-Things (IoT), wearable, and edge systems increasingly require trust in three distinct entities: the human requesting access, the physical device executing the computation, and the hardware function that is permitted to operate. These requirements are usually studied in separate communities. Biometrics establish human identity but remain vulnerable to presentation attacks, intra-user variability, template leakage, and limited revocability. Physical unclonable functions (PUFs) provide device-specific physical identity and on-demand secret derivation, yet must address environmental instability, helper-data exposure, side channels, and modeling attacks. Hardware obfuscation and logic locking condition correct circuit behavior on an activation secret, but face oracle-guided, approximate, structural, removal, and physical attacks.

This survey develops a unified \emph{human--device--function} view of trust. We first decompose each primitive into its complete processing chain and identify the corresponding security assumptions, implementation mechanisms, and evaluation metrics. We then formalize pairwise compositions---biometric--PUF, PUF--obfuscation, and biometric--obfuscation---and a three-way architecture in which correct functionality is bound jointly to an authorized user and a genuine device. Particular attention is given to biometric key reconstruction, PUF stabilization and modeling resistance, logic-locking attack evaluation, cross-layer error propagation, enrollment trust, key lifecycle, and interface leakage. The survey concludes with a taxonomy and research agenda for revocable human--device credentials, compositional security, leakage-aware integration, reconfigurable activation, and standardized end-to-end evaluation.
\end{abstract}


\section{Introduction}
\label{sec:intro}
Authentication in embedded, cyber--physical, and edge systems is increasingly a problem of establishing trust across multiple entities rather than verifying a single credential. A secure system may need to determine not only \emph{who} is requesting access, but also \emph{which physical device} is participating and \emph{whether the requested hardware function is authorized to execute}. These questions correspond to three complementary roots of trust: biometrics can establish trust in the human, physical unclonable functions (PUFs)~\cite{herder2014tutorial,suh2007puf} can establish trust in the device, and hardware obfuscation or logic locking~\cite{forte2017hardware} can enforce trust in the functionality being activated. Although each primitive has been studied extensively in isolation, their composition remains less systematically understood.

The distinction is important because security does not compose automatically. A legitimate biometric may be presented to a counterfeit or compromised device; a genuine device may be operated by an unauthorized user; and a correctly locked circuit may still be activated if its key is exposed, reconstructed, or bypassed. Moreover, once these mechanisms are connected, the interfaces among them become part of the attack surface. A noisy biometric representation can alter a PUF challenge~\cite{karimian2019unlock, shomaji2021blocker, karimian2018secure}, an unstable or modeled PUF response can corrupt a derived activation key, and leakage from the key-derivation or hardware-activation path can bypass otherwise secure upstream mechanisms. The relevant security question is therefore not whether each primitive is secure independently, but whether the complete human--device--function trust chain remains secure under joint operational and adversarial conditions.

This compositional perspective also exposes a common design requirement across the three domains: security-relevant physical or behavioral information must ultimately be transformed into stable, discriminative, and protected digital state. In biometric key generation, noisy physiological or behavioral measurements must be converted into reproducible binary material while preserving inter-user separation and entropy \cite{dodis2008fuzzy, dodis2004fuzzy,uzun2021cryptographic,karimian2019unlock,merhav2018ensemble, karimian2016highly}. PUFs face an analogous problem at the device level, where manufacturing-induced variation must yield responses that are sufficiently unique across devices yet reproducible across temperature, voltage, aging, and measurement noise \cite{tehranipoor2018dvft, tehranipoor2017investigation,suh2007puf,herder2014tutorial}. Memory-based PUFs can exploit existing SRAM or DRAM structures as entropy sources \cite{holcomb2009sram,tehranipoor2017dram,tehranipoor2015dram, yue2020dram}, while challenge--response PUFs must additionally resist prediction from observed CRPs \cite{ruhrmair2010modeling}. At the functional layer, logic locking~\cite{yasin2017evolution} transforms a circuit into a key-dependent implementation whose correct behavior should be recovered only under authorized activation. The SAT attack~\cite{azar2020nngsat,zhou2017cycsat,yasin2016sarlock} demonstrated that nominal key size alone is not a meaningful security guarantee, motivating evaluation in terms of oracle resistance, output corruption, structural leakage, approximate recovery, and physical key exposure \cite{subramanyan2015sat,yasin2016sarlock,xie2019antisat,shamsi2017appsat}.

These observations motivate a unified view of authorization in which human identity, device authenticity, and functional permission are treated as distinct but interacting trust conditions. We express this architectural objective as

\begin{equation}
\label{eq:authorization}
\mathcal{A}
=
\mathcal{T}_{H}
\land
\mathcal{T}_{D}
\land
\mathcal{T}_{F},
\end{equation}

where $\mathcal{T}_{H}$ denotes established trust in the human,
$\mathcal{T}_{D}$ denotes trust in the physical device, and
$\mathcal{T}_{F}$ denotes authorization of the protected function.
Equation~\eqref{eq:authorization} is not intended to imply statistical
independence among these events. In fact, one of the central arguments
of this survey is that their failures can be strongly coupled: errors,
leakage, or compromise at one layer can propagate through key
derivation, device binding, and hardware activation.

Accordingly, this survey studies biometrics, PUFs, and hardware
obfuscation not only as standalone security primitives but as building
blocks of a composed trust architecture. We organize each primitive
around its complete processing pipeline, implementation assumptions,
security properties, failure modes, and evaluation metrics, and then
systematically examine pairwise and three-way compositions. Particular
attention is given to cross-layer reliability, key derivation,
helper-data handling, modeling and side-channel attacks, functional
activation, and end-to-end authorization. The resulting perspective
provides a common framework for evaluating systems in which an
authorized user must interact with an authentic device to activate a
protected hardware function.

%

\FloatBarrier

\section{Human--Device--Function Trust Architecture}
\label{sec:architecture}
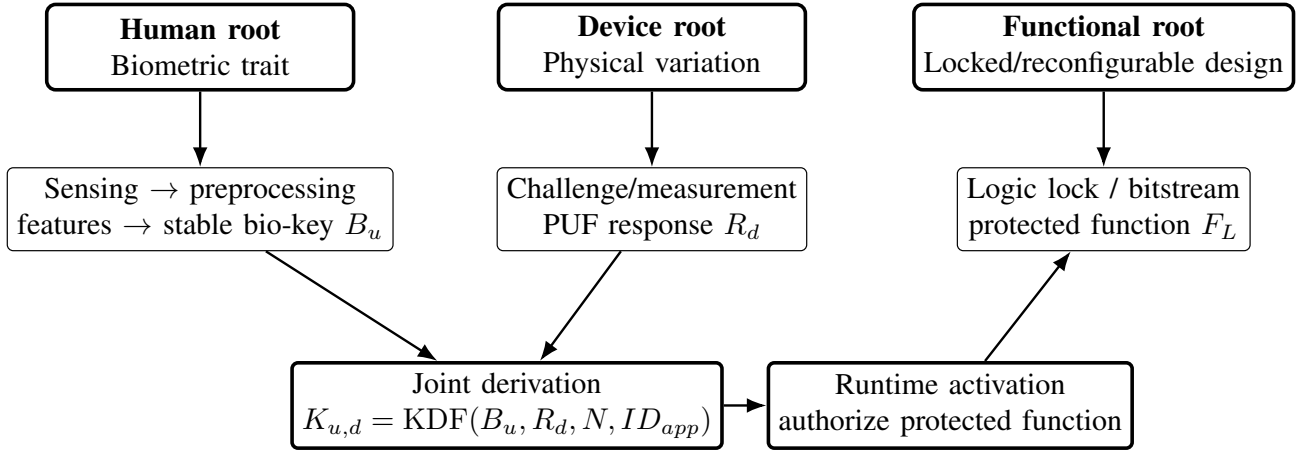
\begin{figure*}[t]
\centering
\resizebox{0.94\textwidth}{!}{%
\begin{tikzpicture}[
    >=Latex,
    root/.style={draw,rounded corners=3pt,very thick,minimum width=3.6cm,minimum height=1.0cm,align=center},
    block/.style={draw,rounded corners=2pt,minimum width=3.1cm,minimum height=0.9cm,align=center},
    fuse/.style={draw,rounded corners=2pt,very thick,minimum width=4.0cm,minimum height=1.0cm,align=center},
    arrow/.style={->,thick}
]
\node[root] (hroot) at (0,3.2) {\textbf{Human root}\\Biometric trait};
\node[root] (droot) at (5.3,3.2) {\textbf{Device root}\\Physical variation};
\node[root] (froot) at (10.6,3.2) {\textbf{Functional root}\\Locked/reconfigurable design};

\node[block] (bio) at (0,1.3) {Sensing $\rightarrow$ preprocessing\\features $\rightarrow$ stable bio-key $B_u$};
\node[block] (puf) at (5.3,1.3) {Challenge/measurement\\PUF response $R_d$};
\node[block] (lock) at (10.6,1.3) {Logic lock / bitstream\\protected function $F_L$};

\node[fuse,minimum width=4.2cm] (kdf) at (3.6,-1.0) {Joint derivation\\$K_{u,d}=\KDF(B_u,R_d,N,ID_{app})$};
\node[fuse,minimum width=3.7cm] (act) at (8.8,-1.0) {Runtime activation\\authorize protected function};

\draw[arrow] (hroot)--(bio);
\draw[arrow] (droot)--(puf);
\draw[arrow] (froot)--(lock);
\draw[arrow] (bio)--(kdf);
\draw[arrow] (puf)--(kdf);
\draw[arrow] (kdf)--(act);
\draw[arrow] (act)--(lock);
\end{tikzpicture}%
}
\caption{Human--device--function trust architecture. Biometrics provide user-dependent information, the PUF contributes device-dependent physical entropy, and logic locking makes correct functionality dependent on their derived authorization key.}
\label{fig:architecture}
\end{figure*}

Figure~\ref{fig:architecture} presents the survey's organizing architecture and separates the human, device, and functional roots of trust from the runtime fusion and activation path.

A general composition can be written as
\begin{align}
B_u &= Q\!\left(f_{\theta}(x_u);\Theta_u\right), \label{eq:biokey}\\
C_u &= H\!\left(B_u\parallel N\parallel ID_{app}\right), \label{eq:pufchallenge}\\
R_{u,d} &= P_d(C_u,E), \label{eq:pufresp}\\
K_{u,d} &= \KDF\!\left(B_u\parallel R_{u,d}\parallel N\parallel ID_{app}\right), \label{eq:jointkey}\\
y &= F_L(x,K_{u,d}), \label{eq:lockedfun}
\end{align}
where $x_u$ is a live biometric acquisition, $Q$ is a reliability-aware quantizer or key-binding mechanism, $E$ represents environmental conditions, $N$ is a nonce, and $ID_{app}$ provides domain separation. Here, $\KDF(\cdot)$ denotes a cryptographic key-derivation function rather than an ad hoc concatenation or hash. Standard constructions such as HKDF follow an extract-then-expand design and provide explicit context separation through the \texttt{info} input \cite{krawczyk2010cryptographic}; NIST SP~800-56C Rev.~2 provides complementary standardized guidance for deriving keying material from established secrets \cite{barker2017recommendation}. The formulation makes explicit which variables are user-dependent, device-dependent, environmental, and application-specific.

\section{Biometrics as a Human Root of Trust}
\label{sec:biometrics}

\subsection{Complete Biometric Processing Chain}
\label{sec:bio_chain}

\begin{figure*}[t]
    \centering
    \includegraphics[width=\textwidth]{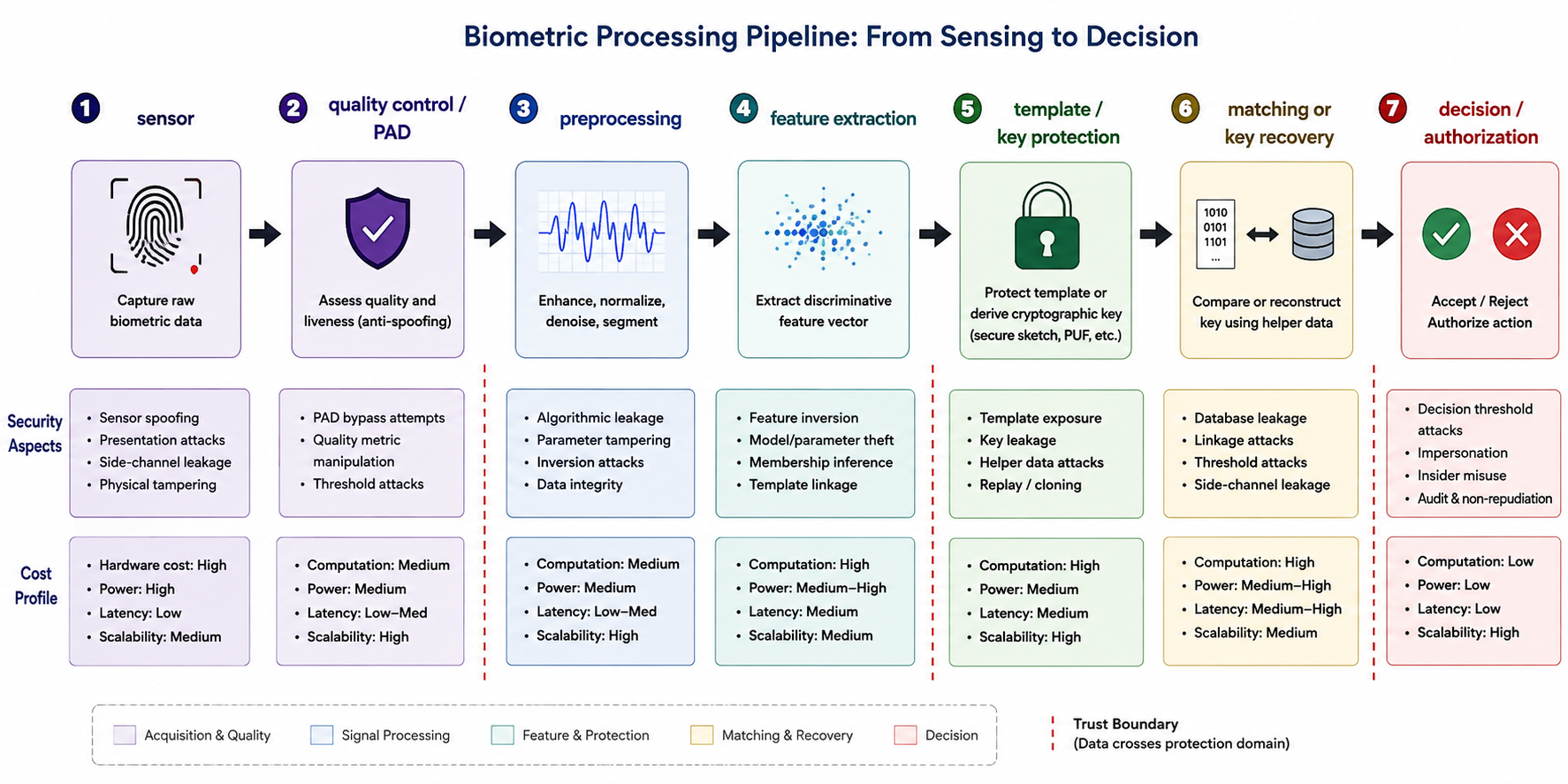}
    \caption{
    End-to-end biometric processing pipeline from sensing to
    authorization. Each stage exposes a distinct attack surface and
    incurs different computational, power, latency, and implementation
    costs. Dashed boundaries indicate transitions between protection
    domains where biometric information changes representation or
    security context.
    }
    \label{fig:biometric_pipeline}
\end{figure*}

A biometric subsystem should be viewed as an end-to-end processing
chain rather than as an isolated matcher. As illustrated in
Fig.~\ref{fig:biometric_pipeline}, biometric information passes through
seven stages: sensing, quality control and presentation-attack
detection (PAD)~\cite{ramachandra2017presentation, george2019biometric}, preprocessing, feature extraction, template or key
protection, matching or key recovery, and finally
decision/authorization. Each stage introduces a different combination
of security exposure, computational cost, power consumption, latency,
and scalability constraints. Consequently, the security of the overall
biometric root of trust depends on the complete pipeline rather than
only on recognition accuracy.

\subsubsection{Sensing}

The biometric sensor forms the first trust boundary because all subsequent
processing depends on the integrity and quality of the acquired signal
\cite{ratha2001enhancing, jain2016fifty, roberts2007biometric, alonso2012quality}.
Depending on the modality, the sensor may capture a fingerprint image~\cite{lin2018matching, sumalatha2024comprehensive}, iris
pattern~\cite{bowyer2008image,wildes2002iris}, facial image~\cite{galbally2014biometric, kim2007method}, speech waveform~\cite{merhav2018ensemble,nautsch2019preserving}, ECG, PPG, or another physiological or
behavioral signal \cite{maltoni2022handbook, daugman2004how, karimian2017highly,
karimian2017nonfiducial,karimian2017human}. At this stage, the system operates on raw biometric
information and is therefore directly exposed to presentation attacks and sensor
spoofing \cite{iso30107part1, marcel2023handbook, matsumoto2002impact,
karimian2017vulnerability}, as well as to physical tampering and
acquisition-channel leakage \cite{uludag2004attacks, tehranipoor2010survey,
zhou2024printlistener, jeon2021gpu, ghandali2021profiled}.

The sensing cost is predominantly determined by the acquisition
hardware rather than by arithmetic computation. Image-based modalities
require optical sensing and, in some cases, controlled illumination or
alignment. Voice and face recognition can often reuse commodity
microphones and cameras. ECG and PPG require dedicated physiological
sensors, although these sensors may already exist in wearable and
medical platforms for health monitoring. This reuse can reduce the
incremental cost of biometric authentication in IoT~\cite{karimian2016evolving, karimian2023never,guo2016hardware} and IoMT~\cite{vishnu2020internet, sun2019security, karimian2019cardiovascular} systems.

Sensor power consumption may dominate the early stage of the pipeline,
particularly for continuously active cameras, high-resolution imaging
systems, or physiological monitoring devices. Latency is generally low
for individual acquisition operations but can increase when the system
requires multiple frames, multiple heartbeats, or repeated samples to
obtain sufficient signal quality. From a security perspective, no
subsequent cryptographic or hardware mechanism can compensate for a
sensor that accepts a successfully spoofed biometric as genuine.

\subsubsection{Quality Control and Presentation-Attack Detection}

Raw acquisition is followed by quality assessment and, where required,
presentation-attack detection. Quality control determines whether the
captured signal contains sufficient information for reliable
downstream processing. Typical checks include signal-to-noise ratio,
image sharpness, region coverage, motion artifacts, illumination,
physiological plausibility, or signal completeness.

PAD extends this stage by attempting to distinguish bona fide
biometric measurements from artificial or replayed traits. Examples
include fingerprint liveness analysis, face anti-spoofing, iris
presentation-attack detection, replay detection for voice, and
physiological consistency tests for ECG or PPG~\cite{karimian2017vulnerability, karimian2020ecg, eberz2017broken, karimian2019attack,garg2021ecg}.

If $q(\mathbf{x})$ denotes a quality function applied to biometric
sample $\mathbf{x}$, the sample may be admitted to the next stage only
when $q(\mathbf{x}) \geq \tau_q$, where $\tau_q$ is a minimum quality threshold. A PAD module can
similarly produce a liveness score

\begin{equation}
\ell(\mathbf{x})
\mathop{\gtrless}_{\mathrm{attack}}^{\mathrm{bona\ fide}}
\tau_{\mathrm{PAD}}.
\label{eq:pad_decision}
\end{equation}

This stage therefore creates an early rejection mechanism before
expensive feature extraction or key reconstruction is performed.

The cost of PAD is strongly implementation-dependent. Lightweight
signal-quality checks may require only simple statistical operations,
whereas learned anti-spoofing systems may introduce a neural network
and become comparable in cost to the biometric recognizer itself.
Security also depends on the robustness of the quality and PAD
thresholds. Manipulation of these thresholds or adversarial
perturbation of quality features can permit malicious inputs to enter
the remainder of the pipeline.

\subsubsection{Preprocessing}

Preprocessing transforms the accepted raw measurement into a more
stable representation by removing nuisance variability unrelated to
identity. Depending on the modality, this stage can include filtering,
normalization, alignment, segmentation, denoising, baseline correction,
artifact removal, image enhancement, or region-of-interest extraction.

Let the acquired biometric signal be $\mathbf{x}$. Preprocessing can
be represented as $\tilde{\mathbf{x}} = g(\mathbf{x};\theta_g)$, where $g(\cdot)$ denotes the preprocessing pipeline and $\theta_g$
contains its associated parameters.

The computational complexity depends heavily on the modality.
For a one-dimensional signal of length $N$, an FIR filter with $L$
coefficients requires approximately $C_{\mathrm{FIR}} = O(NL)$ multiply--accumulate operations. Image-based alignment, enhancement,
and segmentation may require substantially more computation,
particularly when geometric transformations or learned preprocessing
models are used.

Preprocessing is also security relevant. Parameters controlling
filtering, normalization, segmentation, or alignment may be modified to
degrade recognition or bias feature extraction. Timing and power
behavior may reveal information about the input or the processing path.
The timing-aware ECG architecture in
\cite{cordeiro2020timing} demonstrates that different stages of a
biometric algorithm can exhibit distinguishable execution profiles,
creating a side-channel surface even before the final matching stage.

The NA-IOMBA study further illustrates the relationship between
preprocessing and implementation cost. By incorporating biometric
noise into feature-selection and quantization decisions, unnecessary
filtering operations could be reduced, substantially lowering FPGA
resource utilization and power consumption
\cite{karimian2019unlock}. Thus, preprocessing should not be treated
as a fixed front-end operation; it can be co-designed with the
subsequent reliability mechanism.

\subsubsection{Feature Extraction}

Feature extraction converts the preprocessed biometric measurement
into a representation intended to retain identity-related information
while suppressing irrelevant variability. The transformation can be
expressed as $\mathbf{z} = f_{\theta} \left( \tilde{\mathbf{x}} \right)$, where $\mathbf{z} = [z_1,z_2,\ldots,z_m]$ is the resulting $m$-dimensional feature vector.

Classical biometric systems rely on modality-specific handcrafted
features. Examples include minutiae or Gabor responses for
fingerprints, iris codes, spectral or cepstral descriptors for voice,
and fiducial amplitudes, time intervals, or morphological
characteristics for ECG. Modern biometric systems increasingly use
learned embeddings produced by deep neural networks~\cite{sun2015deepid3, minaee2023biometrics, labati2019deep, bhanu2017deep, thentu2021ecg}.

The choice changes both the computational and security profiles.
Handcrafted extraction often has predictable arithmetic complexity and
small model-storage requirements. Learned embeddings may provide
greater representational capacity but require substantially more
multiply--accumulate operations, parameter storage, memory bandwidth,
and often specialized acceleration.

Feature extraction is also a sensitive security boundary because the
feature vector may retain information sufficient to reconstruct or
infer properties of the original biometric. Relevant attacks include
feature inversion, template reconstruction, membership inference,
model extraction, and adversarial manipulation. The implementation may
also reveal information through timing, power, cache, or
electromagnetic side channels.

Thus, feature quality should not be evaluated solely through
recognition performance. A useful feature representation should also
be compact, computationally feasible, stable across repeated
acquisitions, discriminative across users, and sufficiently protected
against inversion or unintended information disclosure.

\subsubsection{Template Protection and Biometric Key Derivation}

The output of feature extraction should generally not be stored in raw
form. Unlike passwords, physiological and behavioral characteristics
cannot be freely replaced following compromise. Template protection
therefore aims to transform $\mathbf{z}$ into a representation that
supports authentication while limiting the consequences of leakage.

A protected representation can be written abstractly as $T_u = \mathcal{P} \left( \mathbf{z}_u; \theta_u,s_u \right)$, where $\mathcal{P}(\cdot)$ denotes a protection transformation, $\theta_u$ contains system- or user-specific parameters, and $s_u$ may provide application-specific diversification or revocation
information.

Template-protection systems are commonly evaluated according to
irreversibility, unlinkability, and renewability or revocability
\cite{jain2008template,nandakumar2015template,patel2015cancelable}.
These properties are also reflected in ISO/IEC~24745:2022, which
specifies requirements and recommendations for protecting biometric
information, including confidentiality, integrity, and renewability/revocability, while ISO/IEC~30136:2018 defines performance
testing and reporting for biometric template-protection schemes, including accuracy, attack success, information leakage, diversity, and
unlinkability \cite{hahn2022biometric, yang2023cross}.
Representative approaches include cancelable transformations, fuzzy
commitment, fuzzy vaults, secure sketches, fuzzy extractors, and
biometric quantization
\cite{juels1999fuzzy,dodis2004fuzzy,karimian2019unlock,hosseinzadehketilateh2024human}.

Biometric key generation imposes a stricter requirement than
conventional template matching. A similarity-based verifier may accept
two nonidentical feature vectors when their distance is below a
threshold. Cryptographic processing, in contrast, normally requires
the same binary key to be recovered exactly. Let $K_u = Q \left( \mathbf{z}_u \right)$ denote a key produced by quantization. For two acquisitions of the same
user, $\mathbf{z}_u^{(1)} \neq \mathbf{z}_u^{(2)}$ is expected because of biometric variability, yet successful
cryptographic reconstruction requires $Q \left( \mathbf{z}_u^{(1)} \right) = Q \left( \mathbf{z}_u^{(2)} \right)$. This requirement motivates reliability-aware feature selection,
adaptive quantization, guard bands, helper data, and error correction.
The computational cost of this stage may therefore include not only
quantization but also reliability estimation, secure-sketch or
fuzzy-extractor operations, error-correcting-code decoding, hashing,
and protected parameter storage.

The security trade-off is similarly nontrivial. Increasing error
tolerance improves reconstruction reliability but may increase helper
data, implementation cost, or information leakage. Selecting only the
most stable features reduces the bit-error rate but can shorten the
key or decrease entropy. Consequently, biometric key-generation
schemes should report reliability, key length, entropy, inter-user
Hamming distance, helper-data overhead, and hardware cost jointly.

\subsubsection{Matching or Key Recovery}

After template protection, the pipeline separates into two closely
related operating modes.

In conventional biometric verification, a probe representation
$\mathbf{z}_p$ is compared with an enrolled representation
$\mathbf{z}_r$ using a similarity function $s = S \left( \mathbf{z}_p,\mathbf{z}_r \right)$. For a similarity-based matcher, authentication proceeds according to $s \mathop{\gtrless}_{\mathrm{reject}}^{\mathrm{accept}} \tau$, where $\tau$ controls the operating trade-off between false acceptance
and false rejection.

In a biometric cryptosystem or key-generation architecture, this stage
instead attempts to reconstruct the enrollment key. If
$\tilde{K}_u$ is the noisy binary~\cite{karimian2017noise} sequence obtained from the current
biometric measurement and $W_u$ denotes available helper data, then $\hat{K}_u = \operatorname{Rep} \left( \tilde{K}_u,W_u \right)$ represents the recovery operation.

The computational cost of conventional matching depends on feature
dimensionality, database size, and matching algorithm. Identification
against a database of $N$ templates may require approximately $C_{\mathrm{match}} = O(Nm)$ operations for direct comparison of $m$-dimensional vectors, unless
indexing or approximate search is used. Key recovery avoids a
large-scale template search but may incur the cost of error correction,
helper-data processing, hashing, or cryptographic verification.

This stage is a prominent attack target because it operates directly
on enrolled and probe representations. Relevant threats include
database leakage~\cite{ignatenko2009biometric}, linkage across applications, manipulation of
similarity scores, hill-climbing attacks~\cite{maiorana2014hill}, helper-data manipulation,
and side-channel leakage from the recovery algorithm.

\subsubsection{Decision and Authorization}

The final stage converts the biometric result into a security action.
In a conventional verifier, the decision variable can be represented
as

\begin{equation}
d_u
=
\begin{cases}
1, & s\geq\tau,\\
0, & s<\tau,
\end{cases}
\label{eq:bio_decision}
\end{equation}

where $d_u=1$ grants authentication.

For biometric key generation, the decision may instead depend on
successful verification of the reconstructed key. If the enrollment
system stores $V_u = H(K_u)$, the current measurement is accepted when $H(\hat{K}_u) = V_u$. In hardware-rooted architectures, authorization can extend beyond a
software access-control decision. The reconstructed key may unlock a
hardware function, generate a PUF challenge, derive a cryptographic
session key, or configure an obfuscated circuit. Thus, the output of
the biometric pipeline can become an input to another root of trust
rather than terminating at a simple Boolean match decision.

The final stage has relatively low arithmetic cost compared with
feature extraction or key recovery, but it is security critical.
Decision overriding, threshold manipulation, privilege escalation,
insider misuse, and improper policy enforcement can nullify the
security of all preceding stages. Auditability and non-repudiation
therefore become relevant once biometric verification is translated
into system-level authorization.

\subsubsection{End-to-End Security and Cost Implications}

Fig.~\ref{fig:biometric_pipeline} emphasizes that biometric security
must be evaluated across the entire processing path. An accurate
matcher does not provide a trustworthy biometric root if the sensor can
be spoofed, the feature representation can be inverted, the helper data
can be manipulated, or the final authorization signal can be overridden.


The total authentication latency can be modeled as

\begin{equation}
T_{\mathrm{bio}}
=
T_{\mathrm{sens}}
+
T_{\mathrm{PAD}}
+
T_{\mathrm{pre}}
+
T_{\mathrm{feat}}
+
T_{\mathrm{prot}}
+
T_{\mathrm{match/rec}}
+
T_{\mathrm{dec}},
\label{eq:total_bio_latency}
\end{equation}

while the corresponding energy cost is

\begin{equation}
\begin{aligned}
E_{\mathrm{bio}}
={}&
E_{\mathrm{sens}}
+E_{\mathrm{PAD}}
+E_{\mathrm{pre}}
+E_{\mathrm{feat}}
\\
&+
E_{\mathrm{prot}}
+E_{\mathrm{match/rec}}
+E_{\mathrm{dec}}.
\end{aligned}
\label{eq:total_bio_energy}
\end{equation}

The terms are not equally dominant across applications. Sensor and
acquisition cost may dominate some physiological or image-based
systems; deep feature extraction may dominate learned recognition
pipelines; error correction or secure key recovery may become
significant in resource-constrained biometric cryptosystems. The same
stage can also present different costs during enrollment and
authentication.

A complete evaluation should therefore report not only biometric
accuracy but also the security and implementation characteristics of
the complete chain. At minimum, this includes FAR, FRR, and EER for
recognition; key reliability, entropy, key length, and reconstruction
rate for biometric cryptosystems; PAD performance where applicable;
and latency, energy, memory, storage, and hardware utilization for the
implementation. This end-to-end view is particularly important in the
human--device--function architecture studied in this survey because
errors or leakage at the biometric layer propagate directly into PUF
challenge generation and hardware authorization.

\subsection{Biometric Decision and Key Metrics}
\label{sec:bio_metrics}

Biometric verification and biometric key generation require
different classes of evaluation metrics. Verification metrics
characterize the overlap between genuine and impostor score
distributions, whereas key-generation metrics quantify whether
a noisy biometric can reproducibly generate sufficiently
unpredictable binary material.

\subsubsection{Decision Metrics}

Let $S$ denote the comparison score, with larger values indicating
greater similarity. Let $p_g(s)$ and $p_i(s)$ denote the score
distributions produced by genuine and impostor comparisons,
respectively. For decision threshold $\tau$, a verification system
accepts a comparison when $S\geq\tau$.

The false acceptance rate is

\begin{equation}
\mathrm{FAR}(\tau)
=
\Pr(S\geq\tau \mid \mathcal{I})
=
\int_{\tau}^{\infty}p_i(s)\,ds,
\label{eq:far}
\end{equation}

where $\mathcal{I}$ denotes an impostor comparison. FAR therefore
measures the probability that an unauthorized subject is incorrectly
accepted.

The false rejection rate is

\begin{equation}
\mathrm{FRR}(\tau)
=
\Pr(S<\tau \mid \mathcal{G})
=
\int_{-\infty}^{\tau}p_g(s)\,ds,
\label{eq:frr}
\end{equation}

where $\mathcal{G}$ denotes a genuine comparison. FRR measures the
probability that the legitimate user is denied access.

Changing $\tau$ creates a security--usability trade-off. Increasing
the threshold generally decreases FAR but increases FRR. The equal
error rate is defined at threshold $\tau^\star$ satisfying

\begin{equation}
\mathrm{FAR}(\tau^\star)
=
\mathrm{FRR}(\tau^\star)
=
\mathrm{EER}.
\label{eq:eer}
\end{equation}

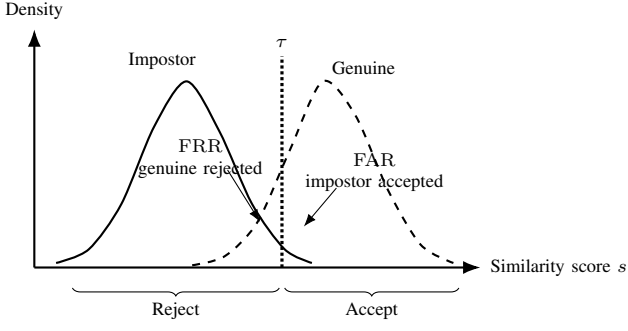
\begin{figure}[t]
\centering
\begin{tikzpicture}[
    x=0.72cm,
    y=3.0cm,
    >=Latex,
    font=\scriptsize
]

\draw[->, thick] (0,0) -- (8.2,0)
    node[right] {Similarity score $s$};

\draw[->, thick] (0,0) -- (0,1.05)
    node[above] {Density};

\draw[thick, smooth]
    plot coordinates {
    (0.4,0.02)
    (1.0,0.08)
    (1.6,0.28)
    (2.2,0.62)
    (2.8,0.82)
    (3.4,0.60)
    (4.0,0.28)
    (4.6,0.08)
    (5.1,0.02)
};

\draw[thick, smooth, dashed]
    plot coordinates {
    (2.9,0.01)
    (3.5,0.05)
    (4.1,0.20)
    (4.7,0.52)
    (5.3,0.82)
    (5.9,0.67)
    (6.5,0.34)
    (7.1,0.10)
    (7.7,0.02)
};

\draw[densely dotted, very thick]
    (4.55,0) -- (4.55,0.93);

\node[above] at (4.55,0.93) {$\tau$};

\node at (2.35,0.91) {Impostor};
\node at (6.05,0.88) {Genuine};

\draw[->] (3.55,0.43) -- (4.15,0.20);
\node[align=center] at (3.05,0.48)
    {$\mathrm{FRR}$\\
     genuine rejected};

\draw[->] (5.55,0.35) -- (4.95,0.18);
\node[align=center] at (6.25,0.42)
    {$\mathrm{FAR}$\\
     impostor accepted};

\draw[decorate,
      decoration={brace,mirror,amplitude=3pt}]
      (0.7,-0.09) -- (4.5,-0.09);

\node at (2.6,-0.19) {Reject};

\draw[decorate,
      decoration={brace,mirror,amplitude=3pt}]
      (4.6,-0.09) -- (7.8,-0.09);

\node at (6.2,-0.19) {Accept};

\end{tikzpicture}

\caption{
Illustration of genuine and impostor biometric score
distributions. For a similarity-based verifier, samples with
$s\geq\tau$ are accepted. The impostor probability mass above
$\tau$ determines the false acceptance rate (FAR), whereas the
genuine probability mass below $\tau$ determines the false
rejection rate (FRR). The equal error rate (EER) is obtained at
the threshold $\tau^\star$ for which $\mathrm{FAR}(\tau^\star)=\mathrm{FRR}(\tau^\star)$. The curves are conceptual and do not represent a particular dataset.
}
\label{fig:biometric_score_distribution}
\end{figure}

Fig.~\ref{fig:biometric_score_distribution} illustrates these
quantities through the overlap of the genuine and impostor score
distributions. EER is convenient for comparing algorithms at a
single operating point, but deployment-oriented studies should also
report FAR and FRR at application-specific thresholds. A system used
for high-security hardware activation may require operation at a
very low FAR even when this produces a higher FRR.

\subsubsection{Biometric Key-Generation Metrics}

Biometric key generation imposes a stricter requirement than
threshold-based recognition. Two measurements of the same person
need not produce identical feature vectors for successful
verification, but cryptographic key reconstruction must reproduce
the required bits exactly or invoke an error-correction mechanism.

For an enrollment key $K_u^{(0)}$ and a reconstructed key
$K_u^{(t)}$ of length $L_K$, key reliability is

\begin{equation}
\mathrm{Rel}_{K}
=
1-
\frac{
\mathrm{HD}
\left(
K_u^{(0)},K_u^{(t)}
\right)
}{
L_K
}.
\label{eq:key_reliability}
\end{equation}

Equivalently, the intra-user bit-error rate is

\begin{equation}
\mathrm{BER}_{\mathrm{intra}}
=
\frac{
\mathrm{HD}
\left(
K_u^{(0)},K_u^{(t)}
\right)
}{
L_K
},
\end{equation}

such that $\mathrm{Rel}_{K} = 1-\mathrm{BER}_{\mathrm{intra}}$. Inter-user separation should also be measured. For keys generated
from different users $u$ and $v$,

\begin{equation}
\mathrm{HD}_{\mathrm{inter}}(u,v)
=
\frac{
\mathrm{HD}(K_u,K_v)
}{
L_K
}.
\label{eq:inter_hd}
\end{equation}

For balanced independent binary keys, the normalized inter-user
Hamming distance is expected to approach $0.5$. Deviation from this
value can indicate correlation, bias, or insufficient user
separation.

The min-entropy of a key variable $K$ is

\begin{equation}
H_{\infty}(K)
=
-\log_2
\left(
\max_k \Pr[K=k]
\right),
\label{eq:minentropy}
\end{equation}

and characterizes the probability of the most likely key value.
For bit-level analysis, min-entropy may also be estimated separately
for individual key positions.

Key length, $L_K = |K|$, must be considered jointly with entropy. A long biometric-derived
binary string does not provide equivalent cryptographic strength if
its bits are strongly biased or correlated.

The probability of successful key reconstruction is another useful
end-to-end metric:

\begin{equation}
P_{\mathrm{KR}}
=
\Pr
\left[
\hat{K}_u = K_u^{(0)}
\right].
\label{eq:key_reconstruction_probability}
\end{equation}

When error-correcting codes or fuzzy extractors are employed,
evaluation should additionally report helper-data size,
error-correction capability, decoding failure probability,
latency, memory, and energy overhead.

\section{Biometric Key Generation Through Quantization}
\label{sec:biometric_key_generation}

Biometric authentication and biometric key generation solve related
but different problems. A conventional biometric verifier can tolerate
moderate intra-user variation because a probe is accepted whenever its
similarity score exceeds a decision threshold. Cryptographic key
generation is less forgiving: measurements obtained from the same user
must repeatedly map to the same binary representation, or the system
must invoke error correction to reconstruct the enrollment key.

Quantization provides a direct mechanism for converting continuous
biometric features into binary key material. Let $\mathbf{x}_u = [x_{u,1},x_{u,2},\ldots,x_{u,d}]$ denote a biometric sample from user $u$. A feature extractor
$f(\cdot)$ produces $\mathbf{z}_u = f(\mathbf{x}_u) = [z_{u,1},z_{u,2},\ldots,z_{u,m}]$, and a quantizer $Q(\cdot)$ maps selected feature values to binary
symbols, $\mathbf{K}_u = Q(\mathbf{z}_u)$. The principal challenge is that biometric measurements are inherently
noisy. Measurements of the same user vary because of sensor noise,
environmental conditions, physiological variation, acquisition
geometry, motion, aging, and preprocessing errors. A quantization
boundary located close to a user's feature distribution can therefore
cause a small perturbation in $z_{u,j}$ to change one or more output
bits.

We distinguish two broad strategies: \emph{fixed quantization} and
\emph{adaptive quantization}. The conceptual adaptive strategy is
illustrated in Fig.~\ref{fig:adaptive_quantization}. In fixed
quantization, all users share predefined intervals. In adaptive
quantization, interval locations, guard bands, or feature-selection
decisions are adapted using enrollment statistics so that unstable
regions can be excluded from key generation.

\subsection{Protocol I: Fixed Quantization}
\label{sec:fixed_quantization}

Fixed quantization maps continuous biometric features into binary
symbols using a common set of quantization boundaries that is shared
across all enrolled users. Unlike adaptive quantization, the location
of these boundaries does not depend on the statistical variability of
a particular subject. The method is therefore simple to implement and
requires little user-specific metadata, but its reliability depends
strongly on how close an individual's biometric features lie to the
predefined decision boundaries.

Let a biometric measurement from user $u$ be denoted by $\mathbf{x}_u$, and let the preprocessing and feature-extraction function
$f(\cdot)$ produce

\begin{equation}
\mathbf{z}_u
=
f(\mathbf{x}_u)
=
\left[
z_{u,1},
z_{u,2},
\ldots,
z_{u,m}
\right],
\label{eq:fixed_feature_vector}
\end{equation}

where $m$ is the number of candidate biometric features.
For feature $j$, assume that its admissible population range is $z_j \in [a_j,b_j]$. A $q$-bit quantizer can represent $M=2^q$ different quantization states. The range
$[a_j,b_j]$ is therefore partitioned into $M$ intervals using the
boundary set $\mathcal{B}_j = \left\{ \beta_{j,0}, \beta_{j,1}, \ldots, \beta_{j,M} \right\}$, with $\beta_{j,0}=a_j, \qquad \beta_{j,M}=b_j$. For uniform fixed quantization, every interval has the same width, $\Delta_j = \frac{b_j-a_j}{M}$, and the corresponding boundaries are

\begin{equation}
\beta_{j,r}
=
a_j+r\Delta_j
=
a_j+
r\frac{b_j-a_j}{M},
\qquad
r=0,\ldots,M.
\label{eq:fixed_boundaries}
\end{equation}

The quantizer determines the region containing the observed feature
value. Specifically, $z_{u,j}$ is assigned to region $r$ when $\beta_{j,r} \leq z_{u,j} < \beta_{j,r+1}$. The region index is then mapped to a binary word, $\mathbf{k}_{u,j} = Q_j(z_{u,j}) = \operatorname{bin}_{q}(r)$, where $\operatorname{bin}_{q}(r)$ denotes the $q$-bit binary
representation assigned to interval $r$.

For example, a two-bit quantizer contains four regions, $M=4$, which may be labeled $00,\quad 01,\quad 10,\quad 11$. As illustrated by the fixed-quantization conceptual plot, every
measurement is assigned to one of these intervals. There is no
explicit exclusion region around a boundary. Consequently, two
measurements that differ only slightly can generate different binary
codes when they fall on opposite sides of the same boundary.

\subsubsection{Enrollment}
\label{sec:fixed_enrollment}

The objective of fixed-quantization enrollment is to construct a
reference biometric key using a globally defined feature-to-bit
mapping. Suppose that $n_e$ biometric measurements are collected from
user $u$,

\begin{equation}
\mathcal{X}_u^{\mathrm{enr}}
=
\left\{
\mathbf{x}_u^{(1)},
\mathbf{x}_u^{(2)},
\ldots,
\mathbf{x}_u^{(n_e)}
\right\}.
\label{eq:fixed_enrollment_set}
\end{equation}

Each acquisition is processed using the same preprocessing and
feature-extraction function, $\mathbf{z}_u^{(t)} = f \left( \mathbf{x}_u^{(t)} \right), \qquad t=1,\ldots,n_e$. When several enrollment measurements are available, they can be
combined into a representative feature vector. A simple choice is the
sample mean, $\bar{z}_{u,j} = \frac{1}{n_e} \sum_{t=1}^{n_e} z_{u,j}^{(t)}$, yielding

\begin{equation}
\bar{\mathbf{z}}_u
=
\left[
\bar{z}_{u,1},
\bar{z}_{u,2},
\ldots,
\bar{z}_{u,m}
\right].
\end{equation}

The important distinction from adaptive quantization is that these
user-specific enrollment statistics do not modify the quantization
boundaries. The same boundary set $\mathcal{B}_j$ is applied to every
user. Thus, the enrollment symbol for feature $j$ is $\mathbf{k}_{u,j}^{(0)} = Q_j \left( \bar{z}_{u,j} \right)$. The complete enrollment key is formed by concatenating the symbols
from all retained features,

\begin{equation}
K_u^{(0)}
=
\mathbf{k}_{u,1}^{(0)}
\Vert
\mathbf{k}_{u,2}^{(0)}
\Vert
\cdots
\Vert
\mathbf{k}_{u,m}^{(0)},
\label{eq:fixed_enrollment_key}
\end{equation}

where $\Vert$ denotes concatenation.

If every feature contributes exactly $q$ bits, the resulting key
length is fixed, $L_K = m q$. This fixed-length property is operationally convenient because the
same processing path, memory allocation, and downstream
cryptographic interface can be used for every user. It also means,
however, that features are typically retained even when their
individual reliability differs substantially across subjects.

The raw biometric key should not generally be stored directly.
Instead, a protected verification value may be generated as

\begin{equation}
V_u
=
H
\left(
K_u^{(0)}
\parallel
\mathrm{ID}_{\mathrm{app}}
\parallel
s_u
\right),
\label{eq:fixed_enrollment_hash}
\end{equation}

where $H(\cdot)$ denotes a cryptographic hash function, $\mathrm{ID}_{\mathrm{app}}$ provides domain separation, and $s_u$ can represent optional diversification or revocation information.

If the expected intra-user variation is too large for exact
reconstruction, helper information can also be derived, $W_u = \operatorname{Gen} \left( K_u^{(0)} \right)$, where $\operatorname{Gen}(\cdot)$ denotes an error-correcting,
secure-sketch, or fuzzy-extractor enrollment function.

The fixed-quantization enrollment pipeline can therefore be summarized
as

\begin{equation}
\boxed{
\mathbf{x}_u
\rightarrow
f(\cdot)
\rightarrow
\bar{\mathbf{z}}_u
\rightarrow
Q_{\mathrm{fixed}}(\cdot)
\rightarrow
K_u^{(0)}
\rightarrow
V_u
}.
\label{eq:fixed_enrollment_pipeline}
\end{equation}

\subsubsection{Authentication and Key Reconstruction}
\label{sec:fixed_authentication}

During authentication, a new biometric sample $\mathbf{x}_u'$ is acquired and processed using the same feature-extraction pipeline,

\begin{equation}
\mathbf{z}_u'
=
f(\mathbf{x}_u')
=
\left[
z_{u,1}',
z_{u,2}',
\ldots,
z_{u,m}'
\right].
\label{eq:fixed_auth_features}
\end{equation}

Each feature is quantized using exactly the same global boundaries
defined during system design. Hence, $\hat{\mathbf{k}}_{u,j} = Q_j \left( z_{u,j}' \right)$, and the directly reconstructed key is

\begin{equation}
\tilde{K}_u
=
\hat{\mathbf{k}}_{u,1}
\Vert
\hat{\mathbf{k}}_{u,2}
\Vert
\cdots
\Vert
\hat{\mathbf{k}}_{u,m}.
\label{eq:fixed_direct_reconstruction}
\end{equation}

Unlike adaptive quantization, no guard-band test is normally applied.
Every feature is mapped to a quantization interval, even when the
measurement is arbitrarily close to a decision boundary. This property
reduces runtime complexity, but it also makes the reconstructed key
more sensitive to intra-user variation.

Consider a feature whose enrollment value lies in interval $r$, $\beta_{j,r} \leq \bar{z}_{u,j} < \beta_{j,r+1}$. A subsequent measurement can be written as $z_{u,j}' = \bar{z}_{u,j} + \epsilon_{u,j}$, where $\epsilon_{u,j}$ represents measurement noise and physiological
variation. The bit assignment changes whenever the perturbation moves
the observation across either neighboring boundary. For the upper
boundary, this occurs when $\epsilon_{u,j} \geq \beta_{j,r+1} - \bar{z}_{u,j}$, while crossing the lower boundary occurs when $\epsilon_{u,j} < \beta_{j,r} - \bar{z}_{u,j}$. The stability of a fixed-quantized feature therefore depends directly
on its distance from the nearest boundary, $d_{u,j}^{\mathrm{fixed}} = \min_r \left| \bar{z}_{u,j} - \beta_{j,r} \right|$. Small values of $d_{u,j}^{\mathrm{fixed}}$ imply a high probability of
symbol disagreement between enrollment and authentication.

If residual errors are present, helper information $W_u$ can be used
to reconstruct the enrollment key, $\hat{K}_u = \operatorname{Rep} \left( \tilde{K}_u, W_u \right)$, where $\operatorname{Rep}(\cdot)$ denotes the corresponding recovery
function.

Successful reconstruction requires $\hat{K}_u = K_u^{(0)}$. As with adaptive quantization, direct storage of
$K_u^{(0)}$ is unnecessary. Instead, the reconstructed key can be
verified through

\begin{equation}
\hat{V}_u
=
H
\left(
\hat{K}_u
\parallel
\mathrm{ID}_{\mathrm{app}}
\parallel
s_u
\right),
\label{eq:fixed_auth_hash}
\end{equation}

and authentication succeeds when $\hat{V}_u = V_u$. The authentication pipeline can therefore be written compactly as

\begin{equation}
\boxed{
\begin{aligned}
\mathbf{x}_u'
&\rightarrow
f(\cdot)
\rightarrow
\mathbf{z}_u'
\rightarrow
Q_{\mathrm{fixed}}(\cdot)
\\
&\rightarrow
\tilde{K}_u
\rightarrow
\operatorname{Rep}(\tilde{K}_u,W_u)
\rightarrow
\hat{K}_u .
\end{aligned}
}
\label{eq:fixed_authentication_pipeline}
\end{equation}

\subsubsection{Reliability Under Fixed Quantization}

The principal limitation of fixed quantization follows from the fact
that the interval structure is population-defined rather than
user-specific. Suppose the authentication feature is modeled as $Z_{u,j}' \sim \mathcal{N} \left( \mu_{u,j}, \sigma_{u,j}^{2} \right)$. If the enrollment feature belongs to interval $r$, the probability
that the same quantized symbol is reproduced is approximately

\begin{equation}
\rho_{u,j}^{\mathrm{fixed}}
=
\Pr
\left[
\beta_{j,r}
\leq
Z_{u,j}'
<
\beta_{j,r+1}
\right].
\label{eq:fixed_reliability_probability}
\end{equation}

Under the Gaussian assumption, this can be expressed as

\begin{equation}
\rho_{u,j}^{\mathrm{fixed}}
=
\Phi
\left(
\frac{
\beta_{j,r+1}-\mu_{u,j}
}{
\sigma_{u,j}
}
\right)
-
\Phi
\left(
\frac{
\beta_{j,r}-\mu_{u,j}
}{
\sigma_{u,j}
}
\right),
\label{eq:fixed_reliability_gaussian}
\end{equation}

where $\Phi(\cdot)$ denotes the standard normal cumulative
distribution function.

Equation~\eqref{eq:fixed_reliability_gaussian} exposes the main
problem with globally fixed boundaries. Two users with the same
feature variance can have very different reconstruction reliability
depending on the location of their respective means within the same
quantization interval. A user's mean located near the center of an
interval can produce a highly reliable bit, whereas another user's
mean close to the boundary may generate frequent bit flips.

The resulting intra-user bit-error rate can be measured as

\begin{equation}
\mathrm{BER}_{\mathrm{intra}}^{\mathrm{fixed}}
=
\frac{
HD
\left(
K_u^{(0)},
\hat{K}_u
\right)
}{
L_K
},
\label{eq:fixed_intra_ber}
\end{equation}

with corresponding key reliability

\begin{equation}
\mathrm{Rel}_{u}^{\mathrm{fixed}}
=
1-
\mathrm{BER}_{\mathrm{intra}}^{\mathrm{fixed}}.
\label{eq:fixed_key_reliability}
\end{equation}

The exact key-reconstruction probability is

\begin{equation}
P_{\mathrm{KR}}^{\mathrm{fixed}}
=
\Pr
\left[
\hat{K}_u
=
K_u^{(0)}
\right].
\label{eq:fixed_pkr}
\end{equation}

If bit errors are assumed independent with average bit-error
probability $p_b$, then a simple approximation without error
correction is $P_{\mathrm{KR}}^{\mathrm{fixed}} \approx (1-p_b)^{L_K}$. This expression illustrates why biometric key generation is more
demanding than conventional biometric matching. Even a relatively
small per-bit error probability can make exact reconstruction of a
long binary key unlikely.

\subsubsection{Entropy and Inter-User Separation}

Fixed quantization must also be evaluated with respect to the
population distribution of the generated symbols. Let $p_{j,r} = \Pr \left[ Z_j \in [\beta_{j,r},\beta_{j,r+1}) \right]$. The entropy of the quantized feature is

\begin{equation}
H(Q_j)
=
-
\sum_{r=0}^{M-1}
p_{j,r}
\log_2 p_{j,r},
\label{eq:fixed_entropy}
\end{equation}

while its min-entropy is

\begin{equation}
H_{\infty}(Q_j)
=
-
\log_2
\left(
\max_r p_{j,r}
\right).
\label{eq:fixed_min_entropy}
\end{equation}

Uniformly spaced intervals do not necessarily produce uniformly
distributed binary symbols. If the biometric feature distribution is
strongly nonuniform, some intervals may occur much more frequently
than others, reducing effective key entropy even though each feature
nominally contributes $q$ bits.

Inter-user separation can be evaluated using normalized Hamming
distance. For users $u$ and $v$,

\begin{equation}
D_{u,v}^{\mathrm{fixed}}
=
\frac{
HD(K_u^{(0)},K_v^{(0)})
}{
L_K
}.
\label{eq:fixed_interuser_hd}
\end{equation}

For balanced and approximately independent binary keys, the expected
normalized inter-user Hamming distance should approach $0.5$.
However, this metric must be interpreted jointly with entropy because
a mean Hamming distance near $0.5$ does not by itself prove
statistical independence or resistance to key prediction.

\subsubsection{Computation and Storage Cost}

Fixed quantization is computationally attractive because the
quantization boundaries are determined globally and can be stored once
for the entire system. For $m$ features and $M$ intervals per feature,
a direct linear search requires approximately $C_{\mathrm{fixed}} = O(mM)$ boundary comparisons. If ordered boundaries are searched using binary
lookup, the cost becomes $C_{\mathrm{fixed}} = O(m\log M)$. For a small number of quantization intervals, the practical cost is
typically close to linear in the number of features, $C_{\mathrm{fixed}} \approx O(m)$. The global boundary storage requirement is $C_{\mathrm{boundary}} = \sum_{j=1}^{m} (M+1)B_{\beta}$, where $B_{\beta}$ denotes the number of bits used to represent each
boundary. Importantly, this cost is independent of the number of
enrolled users.

User-specific storage can consequently be limited to the protected
verification representation and optional helper data, $C_{\mathrm{user}} = |V_u| + |W_u|$. This low user-specific storage and simple online computation make
fixed quantization attractive for resource-constrained hardware.

\subsubsection{Advantages and Limitations}

Fixed quantization offers a deterministic mapping, constant key
length, small user-specific storage, and low enrollment and
authentication complexity. The same quantizer can be implemented for
every user, making hardware realization straightforward.

Its principal limitation is the absence of user-specific reliability
adaptation. The quantizer does not distinguish between a feature
located safely near the center of an interval and one located
arbitrarily close to a boundary. As a consequence, small biometric
variations can produce bit transitions,

\begin{equation}
Q_{\mathrm{fixed}}
\left(
z_{u,j}
\right)
\neq
Q_{\mathrm{fixed}}
\left(
z_{u,j}+\epsilon
\right),
\label{eq:fixed_boundary_flip}
\end{equation}

even when the perturbation $\epsilon$ is small enough that the two
measurements would still be considered the same user by a conventional
biometric matcher.

Fixed quantization therefore tends to shift the burden of robustness
toward preprocessing, filtering, fuzzy extraction, and
error-correction mechanisms. Adaptive quantization takes the opposite
approach: it attempts to avoid generating unreliable bits in the first
place by introducing user-dependent selection criteria and guard
bands, as illustrated in Fig.~\ref{fig:adaptive_quantization}.


\subsection{Protocol II: Adaptive Quantization}
\label{sec:adaptive_quantization}

Adaptive quantization uses enrollment statistics to determine which
regions of the feature space can be mapped reliably to binary symbols.
The central principle is illustrated in
Fig.~\ref{fig:adaptive_quantization}. Colored regions represent
intervals retained for key generation, whereas the regions between
them form \emph{guard bands}. Measurements falling inside a guard
band are treated as unreliable rather than being forced into a
neighboring binary code.

For feature $j$ of user $u$, let enrollment measurements provide the
estimated mean and variability $\mu_{u,j} = \frac{1}{n_e} \sum_{t=1}^{n_e} z_{u,j}^{(t)}$ and

\begin{equation}
\sigma_{u,j}^{2}
=
\frac{1}{n_e-1}
\sum_{t=1}^{n_e}
\left(
z_{u,j}^{(t)}-\mu_{u,j}
\right)^2.
\label{eq:user_variance}
\end{equation}

Rather than applying only global boundaries, the system estimates a
reliability score $\rho_{u,j} = \Pr \left[ Q(z_{u,j}^{(t)}) = Q(z_{u,j}^{(0)}) \right]$. A feature is retained only when $\rho_{u,j} \geq \rho_{\min}$ and when its population-level entropy satisfies $H_j \geq H_{\min}$. The selected feature set is therefore

\begin{equation}
\mathcal{S}_u
=
\left\{
j:
\rho_{u,j}\geq\rho_{\min}
\land
H_j\geq H_{\min}
\right\}.
\label{eq:selected_features}
\end{equation}

This formulation captures the central reliability--entropy trade-off:
features that are highly stable but nearly identical across the
population should not contribute key bits, while highly distinctive
but unstable features should also be excluded.

\subsubsection{Adaptive Guard Bands}

Let $\gamma_{u,j}$ denote the minimum safety margin from a quantization
boundary. A feature measurement is accepted only when $\min_r \left| z_{u,j}-\beta_{j,r} \right| \geq \gamma_{u,j}$. If $\min_r \left| z_{u,j}-\beta_{j,r} \right| < \gamma_{u,j}$, the feature is considered unreliable and no key bit is generated from
that observation.

As illustrated in Fig.~\ref{fig:adaptive_quantization}, this produces
three distinct regions:

\begin{enumerate}[leftmargin=*]

    \item \textbf{selected interval:} the feature is sufficiently far
    from a boundary and is mapped to a binary symbol;

    \item \textbf{guard band:} the measurement is too close to a
    decision boundary and is discarded;

    \item \textbf{unselected feature:} the feature fails the required
    reliability or entropy criterion and is excluded from the key.

\end{enumerate}

\subsubsection{Enrollment}
\label{sec:adaptive_enrollment}

Adaptive quantization differs from fixed quantization primarily in the
way the quantization map is constructed during enrollment. Rather than
applying the same global boundaries to every subject, the enrollment
stage estimates the statistical behavior of each biometric feature for
the individual user and retains only those features that simultaneously
provide sufficient \emph{stability} and \emph{discriminability}. The
result is a user-dependent quantization map in which reliable regions
are assigned binary labels, while unstable regions close to
quantization boundaries are treated as guard bands, as illustrated in
Fig.~\ref{fig:adaptive_quantization}.

For user $u$, assume that $n_e$ biometric measurements are collected
during enrollment,

\begin{equation}
\mathcal{X}_u^{\mathrm{enr}}
=
\left\{
\mathbf{x}_u^{(1)},
\mathbf{x}_u^{(2)},
\ldots,
\mathbf{x}_u^{(n_e)}
\right\}.
\label{eq:adaptive_enrollment_set}
\end{equation}

Each measurement is passed through the same preprocessing and feature
extraction function $f(\cdot)$,

\begin{equation}
\mathbf{z}_u^{(t)}
=
f\!\left(\mathbf{x}_u^{(t)}\right)
=
\left[
z_{u,1}^{(t)},
z_{u,2}^{(t)},
\ldots,
z_{u,m}^{(t)}
\right],
\qquad
t=1,\ldots,n_e,
\label{eq:adaptive_feature_extraction}
\end{equation}

where $m$ denotes the number of candidate biometric features. The
multiple enrollment observations make it possible to estimate the
user-specific distribution of each feature. For feature $j$, its
empirical mean is $\mu_{u,j} = \frac{1}{n_e} \sum_{t=1}^{n_e} z_{u,j}^{(t)}$, and its empirical variance is

\begin{equation}
\sigma_{u,j}^{2}
=
\frac{1}{n_e-1}
\sum_{t=1}^{n_e}
\left(
z_{u,j}^{(t)}-\mu_{u,j}
\right)^2.
\label{eq:user_feature_variance}
\end{equation}

The variance characterizes the intra-user variability of the feature.
A small $\sigma_{u,j}$ indicates that repeated measurements of the
same user remain concentrated, whereas a large value indicates that
the feature is more likely to cross a quantization boundary and
therefore produce unstable key bits.

Reliability alone, however, is not sufficient for key generation. A
feature may be extremely stable for a particular user but have nearly
the same value for most users in the population. Such a feature
contributes little discriminative information or entropy. For this
reason, adaptive quantization combines user-specific stability with
population-level statistics. Let the population distribution of
feature $j$ be represented by the random variable $Z_j$. Its entropy
can be written as

\begin{equation}
H(Z_j)
=
-
\sum_{r}
P(Z_j \in \mathcal{I}_{j,r})
\log_2
P(Z_j \in \mathcal{I}_{j,r}),
\label{eq:feature_entropy}
\end{equation}

where $\mathcal{I}_{j,r}$ denotes the $r$th candidate quantization
interval. For security-sensitive key generation, the corresponding
min-entropy is often more informative,

\begin{equation}
H_{\infty}(Z_j)
=
-
\log_2
\left[
\max_r
P(Z_j \in \mathcal{I}_{j,r})
\right],
\label{eq:feature_min_entropy}
\end{equation}

because it reflects the probability of the most likely quantized
outcome.

The quantization intervals are then chosen so that an enrolled feature
is sufficiently far from the nearest decision boundary. Let
$\beta_{j,r}$ denote a candidate boundary for feature $j$. The
distance between the enrolled feature center and its nearest boundary
is $d_{u,j} = \min_r \left| \mu_{u,j}-\beta_{j,r} \right|$. A reliability margin can be related to the observed feature
variability. For example, one may define a user-specific guard width $\gamma_{u,j} = \kappa\,\sigma_{u,j}$, where $\kappa$ controls the desired tolerance to measurement
variation. The feature is considered suitable for quantization only if

\begin{equation}
d_{u,j}
\geq
\gamma_{u,j}.
\label{eq:guard_condition}
\end{equation}

Equation~\eqref{eq:guard_condition} captures the geometric principle
shown in Fig.~\ref{fig:adaptive_quantization}: feature values located
well inside a colored quantization interval are retained, whereas
values whose expected variation approaches a boundary are excluded by
the guard band.

The same idea can be expressed probabilistically. Let
$Q_{u,j}(\cdot)$ denote the adaptive quantizer for user $u$ and feature
$j$. The reliability of a candidate symbol is

\begin{equation}
\rho_{u,j}
=
\Pr
\left[
Q_{u,j}
\left(
Z_{u,j}^{(t)}
\right)
=
Q_{u,j}
\left(
Z_{u,j}^{(0)}
\right)
\right],
\label{eq:adaptive_reliability}
\end{equation}

where $Z_{u,j}^{(0)}$ represents the enrollment reference and
$Z_{u,j}^{(t)}$ represents a subsequent observation of the same
feature. In practice, this probability is estimated from enrollment
measurements or from a statistical noise model.

A feature is retained only when it satisfies both a reliability
constraint and an entropy constraint. The user-specific selected
feature set can therefore be written as

\begin{equation}
\mathcal{S}_u
=
\left\{
j
\;\middle|\;
\rho_{u,j}
\geq
\rho_{\min}
\;\land\;
H_{\infty}(Z_j)
\geq
H_{\min}
\right\}.
\label{eq:selected_feature_set}
\end{equation}

This selection rule makes explicit the principal trade-off in
biometric key generation. Features with high entropy but poor
reliability are rejected because they generate unstable bits.
Conversely, highly reliable features with insufficient population
entropy are rejected because they are easier to predict and provide
weak inter-user separation.

For each retained feature $j\in\mathcal{S}_u$, the valid feature
space is partitioned into a collection of non-overlapping selected
intervals,

\begin{equation}
\mathcal{I}_{u,j}
=
\left\{
\mathcal{I}_{u,j,0},
\mathcal{I}_{u,j,1},
\ldots,
\mathcal{I}_{u,j,M_j-1}
\right\},
\label{eq:adaptive_intervals}
\end{equation}

where $M_j$ denotes the number of valid quantization states. Each
interval is associated with a binary codeword through a mapping $Q_{u,j}: \mathcal{I}_{u,j,r} \longrightarrow \mathbf{c}_{j,r}$, with

\begin{equation}
\mathbf{c}_{j,r}
\in
\{0,1\}^{q_j},
\qquad
q_j
=
\left\lceil
\log_2 M_j
\right\rceil.
\end{equation}

If the enrollment feature falls inside interval
$\mathcal{I}_{u,j,r}$, the corresponding codeword is $\mathbf{k}_{u,j} = Q_{u,j}(\mu_{u,j}) = \mathbf{c}_{j,r}$. The biometric key is obtained by concatenating the codewords generated
from all retained features, $K_u = \big\Vert_{j\in\mathcal{S}_u} \mathbf{k}_{u,j}$, where $\Vert$ denotes concatenation. Because the selected feature set
and the number of reliable quantization states may differ across
users, the raw biometric key length can also be user dependent, $L_u = |K_u| = \sum_{j\in\mathcal{S}_u} q_j$. This variability is not necessarily a weakness. It reflects the fact
that the method extracts key material only from features for which the
individual user provides sufficient reliability and entropy, rather
than forcing every candidate feature to contribute bits.

The enrollment stage also produces a set of non-secret or
application-dependent reconstruction parameters, denoted collectively
by

\begin{equation}
\Theta_u
=
\left\{
\mathcal{S}_u,
\mathcal{I}_{u,j},
\gamma_{u,j},
q_j,
\ldots
\right\}.
\label{eq:adaptive_parameters}
\end{equation}

Depending on the construction, $\Theta_u$ may be stored directly,
protected as auxiliary data, or incorporated into a secure sketch or
fuzzy-extractor mechanism. If residual bit errors remain after
adaptive feature selection, error-correcting information $W_u$ may
also be generated, $W_u = \operatorname{Gen}(K_u)$, where $\operatorname{Gen}(\cdot)$ represents the helper-data
generation stage of the selected reconstruction mechanism.

The raw biometric key itself should not normally be stored directly.
A protected verification value can instead be computed as $V_u = H \left( K_u \parallel \mathrm{ID}_{\mathrm{app}} \parallel s_u \right)$, where $H(\cdot)$ is a cryptographic hash function,
$\mathrm{ID}_{\mathrm{app}}$ provides application-domain separation,
and $s_u$ denotes optional revocation or diversification data. Thus,
the enrollment output consists of the adaptive quantization parameters
$\Theta_u$, any required helper information $W_u$, and a protected
verification representation $V_u$, rather than a raw biometric
template or plaintext key.

\subsubsection{Authentication and Key Reconstruction}
\label{sec:adaptive_authentication}

During authentication, the objective is to reconstruct the same
enrollment key from a new noisy biometric measurement without
re-estimating the complete user model. Let the authentication sample be $\mathbf{x}_u'$, from which the same preprocessing and feature-extraction pipeline
produces

\begin{equation}
\mathbf{z}_u'
=
f(\mathbf{x}_u')
=
\left[
z_{u,1}',
z_{u,2}',
\ldots,
z_{u,m}'
\right].
\label{eq:authentication_features}
\end{equation}

Only the features selected during enrollment are considered. Thus,
for each $j\in\mathcal{S}_u$, the observed value $z_{u,j}'$ is evaluated using the stored
user-specific quantization parameters $\Theta_u$.

The first step is to determine whether the observation remains inside
a valid selected interval. Define the interval-validity function

\begin{equation}
\delta_{u,j}(z)
=
\begin{cases}
1,
&
z\in
\displaystyle\bigcup_r
\mathcal{I}_{u,j,r},
\\[2mm]
0,
&
\text{otherwise}.
\end{cases}
\label{eq:interval_validity}
\end{equation}

A value for which $\delta_{u,j}(z_{u,j}')=0$ lies outside the selected reliable intervals, typically inside a guard
band or outside the expected enrollment range. Such a measurement is
treated as an erasure rather than being forced into the nearest
quantization state. This distinction is central to adaptive
quantization. Fixed quantization generally assigns every observation
to one side of a boundary; adaptive quantization explicitly admits
that some measurements are too uncertain to produce a trustworthy key
bit.

For a valid feature observation, $\delta_{u,j}(z_{u,j}')=1$, the authentication symbol is $\hat{\mathbf{k}}_{u,j} = Q_{u,j}(z_{u,j}')$. The directly reconstructed key sequence is then

\begin{equation}
\tilde{K}_u
=
\big\Vert_{
j\in\mathcal{S}_u:
\delta_{u,j}(z_{u,j}')=1
}
\hat{\mathbf{k}}_{u,j}.
\label{eq:partial_reconstructed_key}
\end{equation}

Because biometric measurements remain noisy even after adaptive
selection, $\tilde{K}_u$ may contain a small number of bit errors or
erasures. If helper data were generated at enrollment, a
reconstruction function can be applied, $\hat{K}_u = \operatorname{Rep} \left( \tilde{K}_u, W_u \right)$, where $\operatorname{Rep}(\cdot)$ denotes error correction, secure
sketch recovery, or fuzzy-extractor reproduction.

Successful authentication requires reconstruction of the enrollment
key, $\hat{K}_u = K_u$. Because storing $K_u$ directly would undermine the objective of
biometric key protection, equality is normally verified through the
protected enrollment representation. The authentication system
computes

\begin{equation}
\hat{V}_u
=
H
\left(
\hat{K}_u
\parallel
\mathrm{ID}_{\mathrm{app}}
\parallel
s_u
\right)
\label{eq:auth_verification_hash}
\end{equation}

and accepts when $\hat{V}_u = V_u$. The complete adaptive key-generation process can therefore be written
compactly as

\begin{equation}
\boxed{
\mathbf{x}_u
\rightarrow
f(\cdot)
\rightarrow
\mathbf{z}_u
\rightarrow
\mathcal{S}_u
\rightarrow
Q_u(\cdot)
\rightarrow
K_u
\rightarrow
V_u
}
\label{eq:adaptive_enrollment_pipeline}
\end{equation}

during enrollment, and

\begin{equation}
\boxed{
\begin{aligned}
\mathbf{x}_u'
&\rightarrow
f(\cdot)
\rightarrow
\mathbf{z}_u'
\rightarrow
Q_u(\cdot;\Theta_u)
\\
&\rightarrow
\tilde{K}_u
\rightarrow
\operatorname{Rep}(\tilde{K}_u,W_u)
\rightarrow
\hat{K}_u .
\end{aligned}
}
\label{eq:adaptive_authentication_pipeline}
\end{equation}

during authentication.

\begin{figure}[t]
    \centering
    \includegraphics[
        width=\columnwidth
    ]{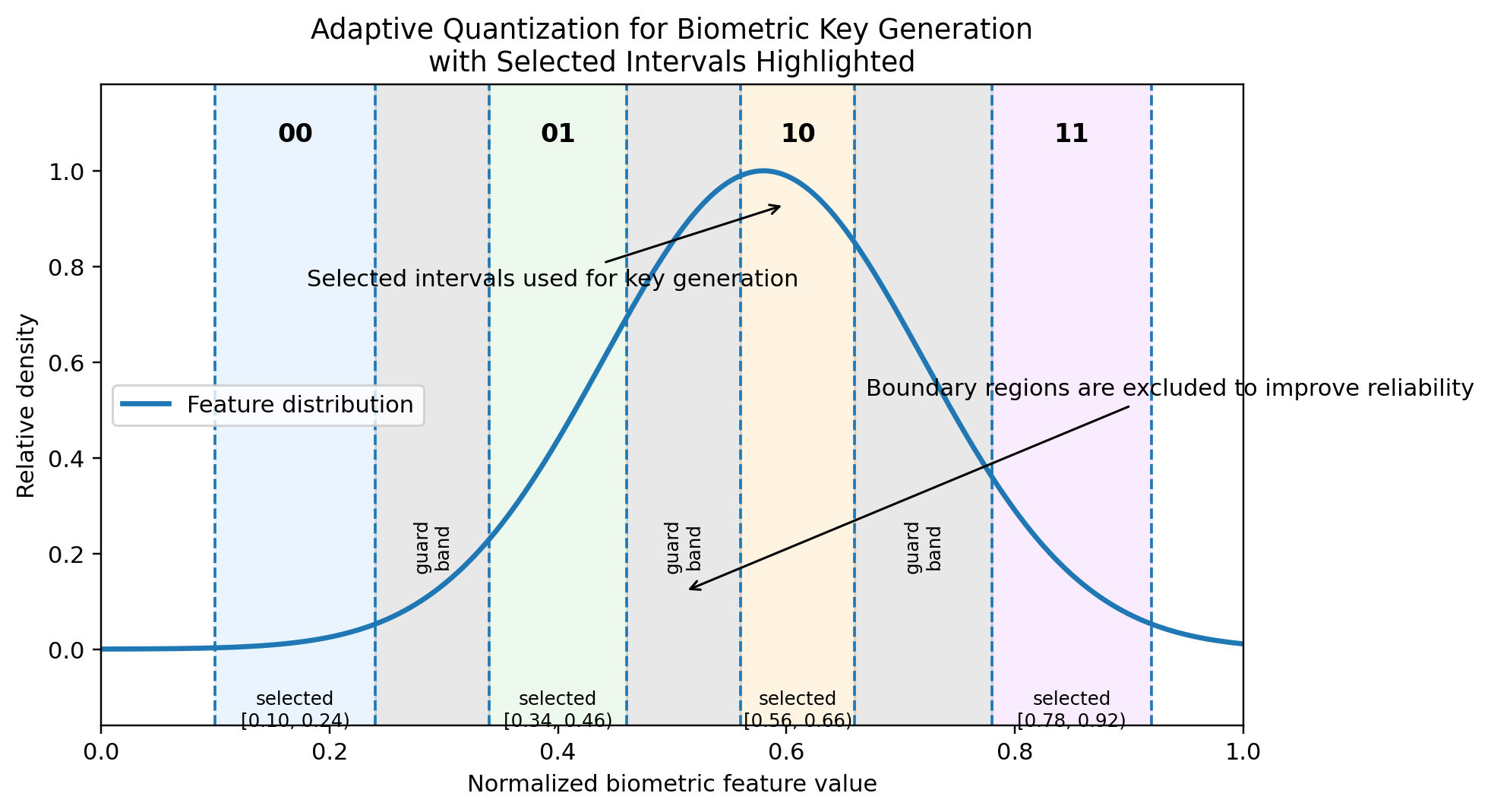}
    \caption{
    Conceptual adaptive quantization for biometric key generation.
    Colored regions represent reliable feature intervals retained for
    bit generation. Gray regions represent guard bands near
    quantization boundaries, where small biometric variations could
    otherwise produce unstable bits. Unlike fixed quantization, the
    adaptive approach can select feature-specific or user-specific
    intervals according to reliability, entropy, and estimated noise.
    The distributions and boundaries are conceptual and are included
    to illustrate the quantization mechanism rather than experimental
    measurements.
    }
    \label{fig:adaptive_quantization}
\end{figure}

The colored intervals in
Fig.~\ref{fig:adaptive_quantization} provide a geometric interpretation
of this process. Only measurements falling in reliable regions
contribute directly to the reconstructed key. Gray guard-band regions
represent feature values for which the probability of a
boundary-crossing error is considered too high. Adaptive quantization
therefore trades raw key length for a lower intra-user bit-error rate
and, consequently, a smaller burden on the subsequent error-correction
mechanism.

For evaluation, the central quantity is the probability of exact key
reconstruction,

\begin{equation}
P_{\mathrm{KR}}
=
\Pr
\left[
\hat{K}_u=K_u
\right].
\label{eq:pkr}
\end{equation}

At the bit level, reconstruction stability can be measured through the
intra-user bit-error rate

\begin{equation}
\mathrm{BER}_{\mathrm{intra}}
=
\frac{
HD(K_u,\hat{K}_u)
}{
|K_u|
},
\label{eq:adaptive_intra_ber}
\end{equation}

with corresponding key reliability $\mathrm{Rel}_u = 1- \mathrm{BER}_{\mathrm{intra}}$. These quantities should be reported jointly with the retained key
length $L_u$, min-entropy $H_{\infty}$, inter-user Hamming distance,
helper-data overhead, and reconstruction cost. A method that achieves
near-perfect reliability by retaining only a few predictable bits is
not suitable for cryptographic key generation; similarly, a method
that produces a long, high-entropy key but reconstructs it
inconsistently is operationally unusable. The purpose of adaptive
quantization is therefore not simply to maximize reliability, but to
identify a useful operating point among \emph{reliability},
\emph{entropy}, \emph{key length}, and \emph{implementation cost}.

\subsection{Fixed Versus Adaptive Quantization:
RQ-Driven Analysis}
\label{sec:fixed_adaptive_analysis}

\subsubsection{RQ1: Which Method Produces More Reliable Keys?}

Fixed quantization forces every measurement into one of the predefined
intervals. A sample located close to a boundary can consequently
change its binary code after a small perturbation.

Adaptive quantization explicitly removes these unstable regions.
Fig.~\ref{fig:adaptive_quantization} illustrates this mechanism:
measurements close to a quantization boundary fall inside guard bands
and are not used directly for key generation.

The principal evaluation metrics are therefore $\mathrm{BER}_{\mathrm{intra}} = \frac{ HD(K_u^{(0)},K_u^{(t)}) }{ |K_u^{(0)}| }$ and $\mathrm{Rel}_{K} = 1-\mathrm{BER}_{\mathrm{intra}}$. Adaptive quantization should primarily be judged by whether the
reduction in intra-user BER justifies the resulting loss in usable
features.

\subsubsection{RQ2: What Is the Reliability--Entropy Trade-off?}

Discarding unstable features improves reliability but reduces the
number of available key bits. Let $L_{\mathrm{raw}}$ denote the number of candidate bits before feature selection and $L_{\mathrm{adapt}} = \left| K_u \right|$ the resulting adaptive key length.

The retention ratio is $\eta_{\mathrm{retain}} = \frac{ L_{\mathrm{adapt}} }{ L_{\mathrm{raw}} }$. A meaningful evaluation should therefore report reliability,
min-entropy, and retained key length simultaneously.

This trade-off is observed in noise-aware biometric quantization:
improving key reliability under adverse conditions may require
sacrificing key length rather than forcing unstable features into the
key \cite{karimian2019unlock}.

\subsubsection{RQ3: Are the Keys Distinct Across Users?}

Reliability alone does not guarantee discriminability. Inter-user
separation can be evaluated as $D_{\mathrm{inter}} = \frac{2}{N(N-1)} \sum_{u<v} \frac{ HD(K_u,K_v) }{ L_K }$. For balanced binary representations, a value near $0.5$ is desirable,
although this statistic alone does not establish cryptographic
independence.

\subsubsection{RQ4: Which Method Is More Robust to Noise?}

Noise sensitivity should be evaluated over a controlled set of
conditions $\mathcal{E} = \{ E_1,E_2,\ldots,E_T \}$, where $E_t$ can represent different SNR levels, sessions, motion
conditions, physiological states, or sensor environments.

Reliability is then reported as $\mathrm{Rel}(E_t) = 1- \frac{ HD(K_u^{(0)},K_u^{(E_t)}) }{ L_K }$. Prior noise-aware ECG quantization results demonstrate the value of
such evaluation: reliability remained high under severe synthesized
noise for NA-IOMBA, whereas conventional IOMBA degraded more strongly
under the same conditions \cite{karimian2019unlock}.

\subsubsection{RQ5: What Is the Computational Cost?}

For $m$ extracted features and $M$ quantization intervals, fixed
quantization requires approximately $C_{\mathrm{fixed}} = O(m\log M)$ comparisons when interval lookup is implemented efficiently.

Adaptive quantization introduces additional enrollment cost:

\begin{equation}
C_{\mathrm{adapt,enroll}}
=
C_{\mathrm{stats}}
+
C_{\mathrm{selection}}
+
C_{\mathrm{interval}}
+
C_{\mathrm{ECC}},
\end{equation}

where the terms correspond to estimating feature statistics,
reliability/entropy selection, interval optimization, and optional
error-correction design.

The online authentication cost can nevertheless be lower because only $m_u = |\mathcal{S}_u|$ selected features need to be processed.

This distinction is important for embedded systems: greater
enrollment complexity can be acceptable when it substantially reduces
runtime preprocessing, error correction, energy, or hardware
utilization.

\subsubsection{RQ6: Which Method Provides the Better System-Level
Trade-off?}

No single metric determines the preferred quantization strategy.
A useful design objective is

\begin{equation}
\begin{aligned}
Q^\star
=
\arg\max_Q
\big[
&
\lambda_1 \mathrm{Rel}
+\lambda_2 H_\infty
+\lambda_3 D_{\mathrm{inter}}
\\
&
-\lambda_4 C_{\mathrm{comp}}
-\lambda_5 C_{\mathrm{HW}}
-\lambda_6 P_{\mathrm{fail}}
\big].
\end{aligned}
\label{eq:quantization_objective}
\end{equation}

where the coefficients $\lambda_i$ represent application-dependent
priorities.

Fixed quantization is attractive when implementation simplicity, small enrollment cost, and deterministic global parameters dominate. Adaptive quantization is preferable when intra-user variability, environmental noise, and error-correction cost are major concerns. For wearable, IoT, and hardware-rooted authentication systems, the latter trade-off is particularly relevant because unreliable biometric bits propagate directly into PUF challenges, cryptographic keys, or hardware-obfuscation keys.

\subsection{Biometric Challenges and Limitations}
Biometric systems face five limitations that become more severe when their outputs directly unlock hardware. First, intra-user variation can convert a tolerable recognition error into complete key failure. Second, raw traits are difficult to revoke, so the protected representation must be renewable. Third, sensor-level spoofing can precede all cryptographic protection. Fourth, feature extraction and matching can leak through timing, power, EM, or memory access. Fifth, the error distribution may be correlated with environment, health, demographics, and device conditions, invalidating simplistic independence assumptions in joint-system analysis.

\FloatBarrier

\section{PUFs as a Device Root of Trust}
\label{sec:pufs}

\subsection{Complete PUF Processing Chain}
\label{sec:puf_chain}

A physical unclonable function (PUF) establishes a hardware-derived
root of trust by exploiting manufacturing variations that are difficult
to reproduce across nominally identical devices. Early silicon PUF
constructions demonstrated that fabrication-induced delay variation
could be used to generate device-dependent challenge--response behavior
without storing the corresponding secret in conventional nonvolatile
memory \cite{gassend2002silicon,suh2007puf}. Since then, PUFs have been
used primarily for device authentication, hardware fingerprinting, and
volatile cryptographic key generation. The design space spans delay,
memory, coating, optical, and emerging device constructions.

A PUF instance implemented on device $d$ can be represented abstractly
as

\begin{equation}
R
=
P_d(C,E,A),
\label{eq:puf_model_complete}
\end{equation}

where $C$ denotes the applied challenge, $R$ denotes the raw physical
response, $E$ captures environmental conditions such as temperature and
supply voltage, and $A$ represents time-dependent effects such as aging.
The function $P_d(\cdot)$ is determined by device-specific physical
variation rather than by a digitally stored secret.

\begin{figure*}[t]
    \centering
    \includegraphics[width=0.98\textwidth]{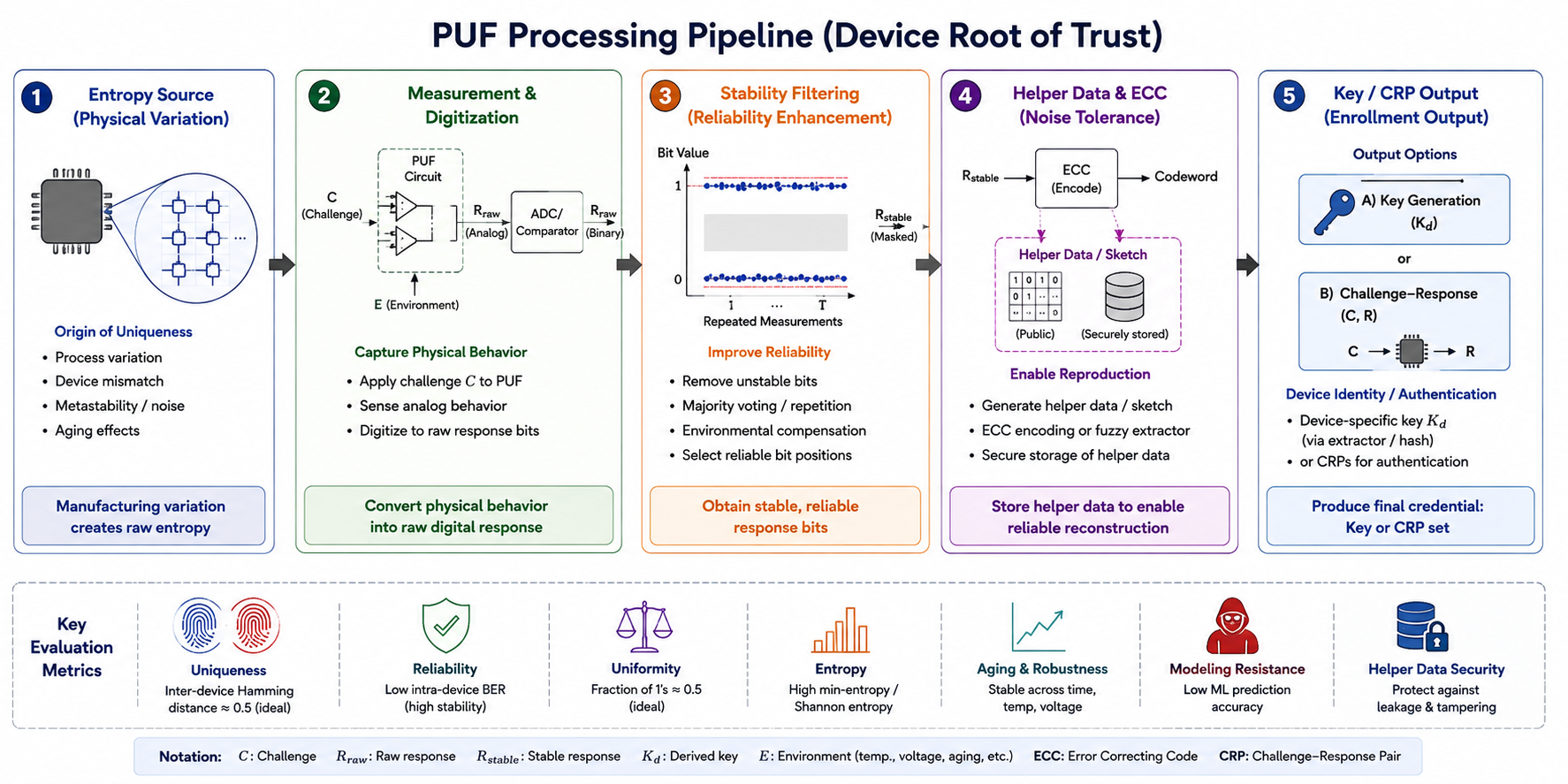}
    \caption{
    \textbf{End-to-end PUF processing pipeline for device-rooted
    authentication and key generation.}
    Manufacturing variation first produces a device-specific physical
    entropy source. The physical behavior is measured and digitized into
    a raw response, after which unstable response positions can be
    removed or stabilized through repeated measurements, reliability
    screening, or environmental characterization. Secure sketches,
    fuzzy extractors, helper data, and error-correcting codes can then be
    used to reproduce a stable secret from noisy measurements. The final
    output is either a device-derived cryptographic key or a
    challenge--response credential used for authentication. The lower
    portion summarizes the principal PUF evaluation metrics and the
    security--cost trade-offs of major PUF families.
    }
    \label{fig:puf_pipeline}
\end{figure*}

Figure~\ref{fig:puf_pipeline} shows the complete processing path from
the physical entropy source to a stable key or challenge--response pair.
The important distinction is that the raw PUF measurement is rarely the
final security credential. Practical systems typically require
digitization, reliability filtering, helper-data generation~\cite{yan2015novel}, error
correction, and cryptographic post-processing before the output is
suitable for authentication or key derivation.

The processing chain can be represented compactly as

\begin{equation}
\boxed{
\mathcal{V}_d
\rightarrow
R_{\mathrm{raw}}
\rightarrow
R_{\mathrm{stable}}
\rightarrow
(K,W)
\rightarrow
K_d \;\text{or}\; (C,R)
}
\label{eq:puf_pipeline_compact}
\end{equation}

where $\mathcal{V}_d$ denotes the physical variation source,
$R_{\mathrm{raw}}$ the directly observed response,
$R_{\mathrm{stable}}$ the reliability-screened response,
$W$ helper information, and $K_d$ the final device-specific key.

\subsubsection{Physical Entropy Source}
\label{sec:puf_entropy_source}

The first stage is the physical mechanism from which device identity is
derived. Depending on the PUF family, the relevant variation may arise
from transistor threshold mismatch, interconnect delay, oscillator
frequency, SRAM power-up preference, DRAM retention behavior, or
stochastic resistance and switching characteristics in emerging
devices.

For two physical devices $d_i$ and $d_j$ fabricated from the same
nominal design, a useful PUF should satisfy $P_{d_i}(C,E) \neq P_{d_j}(C,E), \qquad d_i\neq d_j$, with high probability. At the same time, repeated measurements of the
same physical instance should remain sufficiently close, $P_d(C,E_1) \approx P_d(C,E_2)$. These requirements create the central PUF design tension:
manufacturing variation must be strong enough to distinguish devices
but insensitive enough to environmental perturbation that the same
device can reproduce its response.

The entropy source itself may require little or no dedicated area when
existing structures are reused. SRAM PUFs, for example, exploit the
preferred power-up state of existing SRAM cells
\cite{holcomb2009sram}. DRAM PUFs similarly reuse memory behavior,
including startup, retention, or reduced-latency effects
\cite{tehranipoor2017dram}. In contrast, delay-based PUFs generally
require dedicated delay paths, arbiters, oscillators, or measurement
logic.

\subsubsection{Challenge Application, Measurement, and Digitization}
\label{sec:puf_measurement}

The physical entropy source must be converted into a reproducible
digital observation. Let $Y_{d}(C,E)$ denote the analog physical quantity measured for challenge $C$. The
digitizer maps this quantity into a raw binary response, $R_{\mathrm{raw}} = D \left( Y_d(C,E) \right)$, where $D(\cdot)$ may represent an arbiter decision, frequency
comparison, voltage threshold, SRAM state readout, DRAM failure
observation, or analog sensing operation.

For an arbiter PUF, two nominally symmetric paths compete, and the
response can be written as

\begin{equation}
R(C)
=
\begin{cases}
1, & \Delta(C) > 0,\\
0, & \Delta(C) \leq 0,
\end{cases}
\label{eq:arbiter_response}
\end{equation}

where $\Delta(C)$ is the challenge-dependent delay difference between
the two paths. The original silicon physical random function work
established this delay-based construction and its use for
challenge--response identification \cite{gassend2002silicon}.

For a ring-oscillator PUF, a common response is obtained by comparing
the frequencies of two oscillators,

\begin{equation}
R_{i,j}
=
\begin{cases}
1, & f_i > f_j,\\
0, & f_i \leq f_j.
\end{cases}
\label{eq:ro_response}
\end{equation}

The RO-PUF therefore requires a measurement window long enough to
estimate oscillator frequencies reliably. Improved configurable
RO-PUF designs study this trade-off between response quality,
measurement time, and FPGA implementation overhead
\cite{maiti2010ro}.

Memory-based PUFs use a different digitization mechanism. For SRAM,
the response bit can be modeled as

\begin{equation}
R_{\ell}
=
\begin{cases}
1, & V_{\mathrm{mismatch},\ell}>0,\\
0, & V_{\mathrm{mismatch},\ell}\leq0,
\end{cases}
\label{eq:sram_bit}
\end{equation}

where $V_{\mathrm{mismatch},\ell}$ abstracts the mismatch that biases
cell $\ell$ toward one of its two startup states.

\subsubsection{Stability Filtering and Reliable-Bit Selection}
\label{sec:puf_stability_filter}

Raw PUF responses are not perfectly deterministic. Supply-voltage
variation, temperature, measurement noise, aging, metastability, and
device disturbance can cause individual response bits to flip.
Consequently, a practical system commonly characterizes each response
position during enrollment.

Suppose response bit $\ell$ is measured $T$ times, $\left\{ R_{\ell}^{(1)}, R_{\ell}^{(2)}, \ldots, R_{\ell}^{(T)} \right\}$. Its empirical probability of producing a one is $\hat{p}_{d,\ell} = \frac{1}{T} \sum_{t=1}^{T} R_{\ell}^{(t)}$. A simple bit-confidence measure is $\Gamma_{d,\ell} = \left| 2\hat{p}_{d,\ell}-1 \right|$, where values near one indicate highly stable bits and values near zero
indicate unstable or metastable positions.

A reliable-bit mask can therefore be defined as

\begin{equation}
M_{d,\ell}
=
\begin{cases}
1,
&
\Gamma_{d,\ell}\geq \gamma_{\min},
\\
0,
&
\Gamma_{d,\ell}<\gamma_{\min}.
\end{cases}
\label{eq:puf_stable_mask}
\end{equation}

The filtered response is $R_{\mathrm{stable}} = R_{\mathrm{raw}} \odot M_d$, where $\odot$ indicates selection of positions retained by the mask.

Repeated measurement and majority voting provide another common
stabilization mechanism. For $T$ odd repeated measurements,

\begin{equation}
\hat{R}_{\ell}
=
\mathbb{I}
\left[
\sum_{t=1}^{T}
R_{\ell}^{(t)}
>
\frac{T}{2}
\right].
\label{eq:puf_majority_vote}
\end{equation}

These mechanisms improve reliability but increase enrollment time,
runtime latency, energy, and storage for reliability metadata.

\subsubsection{Helper Data, Secure Sketches, and Fuzzy Extraction}
\label{sec:puf_helper_data}

For key-generation applications, filtering alone is often insufficient. The same device must reproduce the exact enrollment key even when a small number of PUF bits change. Error-correcting codes and fuzzy extractors therefore form a central part of many practical PUF key generators.

Let the enrollment response be $R_0$. A fuzzy extractor can be written
as $(K_d,W) = \operatorname{Gen}(R_0)$, where $K_d$ is the derived secret and $W$ is helper data. At a later
time, a noisy response $R_t$ is processed as $\hat{K}_d = \operatorname{Rep}(R_t,W)$. For an error-correction radius $t_c$, the desired reconstruction
condition is

\begin{equation}
HD(R_0,R_t)
\leq
t_c
\quad
\Longrightarrow
\quad
\hat{K}_d
=
K_d.
\label{eq:puf_fe_correctness}
\end{equation}

If an $[n,k,d_{\min}]$ error-correcting code is used, the conventional
hard-decision correction capability is $t_c = \left\lfloor \frac{d_{\min}-1}{2} \right\rfloor$. The redundancy introduced by the ECC is $r_{\mathrm{ECC}} = n-k$, while the code rate is $R_{\mathrm{code}} = \frac{k}{n}$. Increasing $t_c$ generally requires greater redundancy, larger decoding
logic, more helper data, and higher latency. Error correction therefore
creates a direct reliability--cost trade-off.

Helper data must also be included in the security model. Although
helper data need not reveal the reconstructed secret directly, active
manipulation can alter decoding behavior or reduce effective security.
Becker showed that practical PUF fuzzy extractors require careful
analysis under helper-data manipulation rather than only passive
helper-data exposure \cite{becker2017helper}. More recent PUF
key-generation work has consequently proposed explicit helper-data
masking mechanisms \cite{pour2022helper}.

\subsubsection{Key Derivation and Challenge--Response Output}
\label{sec:puf_key_output}

After stabilization, a PUF may support either local key reconstruction
or remote challenge--response authentication.

For local key generation, the stable response can be passed through a
cryptographic extractor,

\begin{equation}
K_d
=
H
\left(
R_{\mathrm{stable}}
\parallel
s_d
\parallel
ID_{\mathrm{app}}
\right),
\label{eq:puf_key_derivation}
\end{equation}

where $H(\cdot)$ is a cryptographic hash function, $s_d$ is optional
device-specific diversification information, and
$ID_{\mathrm{app}}$ provides application separation.

The security objective is that the key be regenerated when needed but
not persistently stored, $K_d^{(t)} = K_d^{(0)}$ for legitimate repeated measurements after stabilization.

For remote PUF authentication, the verifier instead stores or models a
set of challenge--response relations, $\mathcal{D}_d = \left\{ (C_i,R_i) \right\}_{i=1}^{N_{\mathrm{CRP}}}$. A device is authenticated if its observed response remains within the
accepted distance from the enrolled response, $\frac{ HD(R_i,\hat{R}_i) }{ L } \leq \tau_R$.
\subsection{Weak and Strong PUFs}
\label{sec:weak_strong_puf}

PUFs are commonly distinguished according to the size and accessibility
of their challenge--response space. Weak PUFs expose a limited number
of effectively independent responses and are most naturally used for
local key generation or device fingerprinting. SRAM and many
memory-based PUFs fall into this category because the device exposes a
fixed or relatively small response space.

Strong PUFs provide a much larger challenge space, $|\mathcal{C}| \gg 1$, and are intended to support authentication through many
challenge--response pairs. Arbiter-derived constructions are canonical
examples \cite{gassend2002silicon,tehranipoor2017study,suh2007puf}.

A large CRP space, however, does not itself imply unpredictability.
Rührmair \emph{et al.} demonstrated that several strong PUF
constructions could be approximated using machine-learning models from
observed CRPs \cite{ruhrmair2010modeling}. Later neural-network attacks
substantially extended practical modeling capability against XOR,
feed-forward, and interpose arbiter-based PUF constructions
\cite{wisiol2022neural}. Modeling resistance must therefore be treated
as a first-class evaluation metric rather than inferred from challenge
space alone.


\subsection{PUF Evaluation Metrics}
\label{sec:puf_metrics}

PUF evaluation must characterize both statistical quality and
adversarial predictability. High uniqueness or reliability does not
imply resistance to modeling, and modeling resistance does not
guarantee environmental stability. ISO/IEC~20897-1:2020 provides
security requirements for PUF output properties, tamper resistance,
and unclonability, while ISO/IEC~20897-2:2022 specifies test and
evaluation methods based on design inspection and statistical analysis
of PUF responses according to International Organization for Standardization/ International Electrotechnical Commission. These
standards provide a useful baseline for reporting PUF quality, but
application-specific studies should additionally evaluate modeling,
helper-data, side-channel, and environmental attacks.

\subsubsection{Inter-Device Uniqueness}

For $N$ devices evaluated under the same challenge set and response
length $L$, uniqueness is commonly defined as the mean normalized
inter-device Hamming distance,

\begin{equation}
\mathrm{Uniq}
=
\frac{2}{N(N-1)}
\sum_{i=1}^{N-1}
\sum_{j=i+1}^{N}
\frac{
HD(R_i,R_j)
}{
L
}.
\label{eq:uniqueness}
\end{equation}

For independent balanced binary responses, the ideal value is $\mathrm{Uniq}_{\mathrm{ideal}} = 0.5$. A value substantially below $0.5$ indicates that different devices
produce overly similar responses, while large systematic deviations
can reveal population bias.

\subsubsection{Intra-Device Reliability}

Let $R_i^{(0)}$ be the enrolled response of device $i$ and
$R_i^{(t)}$ its $t$th repeated measurement. The normalized intra-device
bit-error rate is

\begin{equation}
\mathrm{BER}_{i}
=
\frac{1}{T}
\sum_{t=1}^{T}
\frac{
HD
\left(
R_i^{(0)},R_i^{(t)}
\right)
}{
L
}.
\label{eq:puf_ber}
\end{equation}

The corresponding reliability is $\mathrm{Rel}_{i} = 1-\mathrm{BER}_{i}$. A population-wide reliability metric can be written as

\begin{equation}
\mathrm{Rel}
=
1-
\frac{1}{NT}
\sum_{i=1}^{N}
\sum_{t=1}^{T}
\frac{
HD
\left(
R_i^{(0)},R_i^{(t)}
\right)
}{
L
}.
\label{eq:pufrel}
\end{equation}

For key reconstruction, the distribution and worst case are often more
informative than only the mean, $\mathrm{Rel}_{\min} = \min_i \mathrm{Rel}_i$.
\subsubsection{Uniformity}

Uniformity measures the proportion of ones in the response of one
device,

\begin{equation}
\mathrm{Unif}_{i}
=
\frac{1}{L}
\sum_{\ell=1}^{L}
R_i[\ell].
\label{eq:uniformity}
\end{equation}

For balanced binary responses, $\mathrm{Unif}_{\mathrm{ideal}} = 0.5$. Uniformity is useful for detecting response bias but should not be
interpreted as a complete entropy metric.

\subsubsection{Bit Aliasing}

Bit aliasing evaluates population-level bias at response position $\ell$,
\begin{equation}
\mathrm{Alias}[\ell]
=
\frac{1}{N}
\sum_{i=1}^{N}
R_i[\ell].
\label{eq:bit_aliasing}
\end{equation}

Ideally, $\mathrm{Alias}[\ell] \approx 0.5 \qquad \forall \ell$. A response position that produces the same value across most devices
provides little device-specific information even if it is perfectly
stable.

\subsubsection{Response Entropy}

For a binary response bit $R[\ell]$ with probability $p_{\ell} = \Pr[R[\ell]=1]$, its Shannon entropy is

\begin{equation}
H(R[\ell])
=
-
p_{\ell}\log_2 p_{\ell}
-
(1-p_{\ell})
\log_2(1-p_{\ell}).
\label{eq:puf_shannon_entropy}
\end{equation}

The corresponding min-entropy is

\begin{equation}
H_{\infty}(R[\ell])
=
-
\log_2
\left(
\max
\{
p_{\ell},
1-p_{\ell}
\}
\right).
\label{eq:puf_minentropy}
\end{equation}

For a response vector, simply summing individual bit entropies assumes
independence. Correlation among response bits should therefore be
measured separately when claiming cryptographic key entropy.

\subsubsection{Environmental Robustness}

Reliability should be characterized across a set of operating
conditions $\mathcal{E} = \{ E_1,E_2,\ldots,E_M \}$. Condition-dependent reliability can be written as

\begin{equation}
\mathrm{Rel}(E_m)
=
1-
\frac{
HD
\left(
R(E_0),R(E_m)
\right)
}{
L
},
\label{eq:puf_env_rel}
\end{equation}

where $E_0$ denotes the reference enrollment condition.

A useful deployment metric is worst-case reliability,

\begin{equation}
\mathrm{Rel}_{\mathrm{worst}}
=
\min_{E_m\in\mathcal{E}}
\mathrm{Rel}(E_m).
\label{eq:puf_worst_reliability}
\end{equation}

Aging can be represented similarly,

\begin{equation}
\mathrm{BER}_{\mathrm{age}}(t)
=
\frac{
HD
\left(
R(0),R(t)
\right)
}{
L
}.
\label{eq:puf_aging_ber}
\end{equation}

RO-PUF studies have explicitly examined aging-induced reliability loss
and circuit-level mitigation \cite{rahman2015aging}.

\subsection{Modeling-Attack Evaluation}
\label{sec:puf_modeling_metrics}

For a strong PUF, statistical response quality must be supplemented by
an explicit modeling-attack experiment~\cite{ruhrmair2013puf, delvaux2019machine}. Suppose an adversary observes $\mathcal{D}_{\mathrm{train}} = \{ (C_i,R_i) \}_{i=1}^{n}$ and learns a model $\hat{P}_{\phi}(C) \approx P_d(C)$. For binary responses, prediction accuracy on an unseen test set is

\begin{equation}
\mathrm{Acc}_{\mathrm{ML}}
=
\frac{1}{
|\mathcal{D}_{\mathrm{test}}|
}
\sum_{(C,R)\in\mathcal{D}_{\mathrm{test}}}
\mathbb{I}
\left[
\hat{P}_{\phi}(C)=R
\right].
\label{eq:puf_ml_accuracy}
\end{equation}

For balanced binary PUFs, random prediction gives approximately $\mathrm{Acc}_{\mathrm{random}} = 0.5$. A simple modeling advantage can therefore be written as $\mathrm{Adv}_{\mathrm{ML}} = \left| \mathrm{Acc}_{\mathrm{ML}} - 0.5 \right|$. The CRP sample complexity required to reach attack accuracy
$\tau_{\mathrm{atk}}$ can be defined as

\begin{equation}
n_{\mathrm{break}}
=
\min
\left\{
n:
\mathrm{Acc}_{\mathrm{ML}}(n)
\geq
\tau_{\mathrm{atk}}
\right\}.
\label{eq:puf_crps_break}
\end{equation}

Reporting only prediction accuracy without $n_{\mathrm{break}}$ can be
misleading because two PUFs may reach the same attack accuracy but
require very different numbers of exposed CRPs. The original modeling
attack literature established CRP-based machine-learning attacks as a
central threat to strong PUFs \cite{ruhrmair2010modeling}; later
neural-network work showed that several ostensibly harder arbiter-PUF
variants could be modeled with fewer samples and at larger design
parameters than previously demonstrated \cite{wisiol2022neural}.

\subsection{Helper-Data and ECC Overhead}
\label{sec:puf_ecc_cost}

The cost of PUF reliability cannot be assessed from the PUF core alone.
For a raw response of length $L_{\mathrm{raw}}$, let only
$L_{\mathrm{sel}}$ stable bits remain after reliability screening. The
selection efficiency is $\eta_{\mathrm{sel}} = \frac{ L_{\mathrm{sel}} }{ L_{\mathrm{raw}} }$. If an ECC with parameters $[n,k]$ is used, its storage redundancy is $O_{\mathrm{ECC}} = n-k$, and normalized redundancy is $\eta_{\mathrm{ECC}} = \frac{n-k}{k}$. The helper-data storage cost can be expressed as $S_{\mathrm{helper}} = |W|$. The complete PUF key-generator storage requirement becomes

\begin{equation}
S_{\mathrm{PUF}}
=
S_{\mathrm{mask}}
+
S_{\mathrm{helper}}
+
S_{\mathrm{metadata}}
+
S_{\mathrm{hash}},
\label{eq:puf_storage_total}
\end{equation}

where the PUF-derived secret itself need not be stored persistently.

In systems that combine biometrics and PUFs, separate fuzzy extraction
at both layers can become expensive. A biometric subsystem may require
error correction to stabilize $B_u$, while the PUF independently
requires ECC for $R_d$. The total correction latency is then

\begin{equation}
T_{\mathrm{corr,total}}
=
T_{\mathrm{bio\text{-}ECC}}
+
T_{\mathrm{PUF\text{-}ECC}}.
\label{eq:dual_ecc_latency}
\end{equation}

This motivates joint reliability selection before invoking expensive
decoding.

\subsection{PUF Computational and Hardware Cost}
\label{sec:puf_cost}

The computational cost of a PUF depends strongly on whether the
underlying physical structure already exists in the target platform.

For one response evaluation, total latency can be modeled as

\begin{equation}
T_{\mathrm{PUF}}
=
T_{\mathrm{stim}}
+
T_{\mathrm{measure}}
+
T_{\mathrm{digit}}
+
T_{\mathrm{filter}}
+
T_{\mathrm{ECC}}
+
T_{\mathrm{extract}}.
\label{eq:puf_total_latency}
\end{equation}

The corresponding energy cost is

\begin{equation}
E_{\mathrm{PUF}}
=
E_{\mathrm{stim}}
+
E_{\mathrm{measure}}
+
E_{\mathrm{digit}}
+
E_{\mathrm{filter}}
+
E_{\mathrm{ECC}}
+
E_{\mathrm{extract}}.
\label{eq:puf_total_energy}
\end{equation}

Area overhead can be written as

\begin{equation}
A_{\mathrm{PUF}}
=
A_{\mathrm{core}}
+
A_{\mathrm{readout}}
+
A_{\mathrm{control}}
+
A_{\mathrm{ECC}}
+
A_{\mathrm{helper}}.
\label{eq:puf_area}
\end{equation}

For memory PUFs, $A_{\mathrm{core}} \approx 0$ may be a reasonable architectural approximation when existing memory is
reused, although control, stable-bit selection, helper data, and ECC
remain nonzero. SRAM power-up PUFs exemplify this reuse model
\cite{holcomb2009sram}.

DRAM PUFs~\cite{anagnostopoulos2018securing,anagnostopoulos2018overview} likewise reuse commodity memory but often incur substantially
larger measurement latency because refresh behavior, retention windows,
reduced timing parameters, or repeated memory accesses must be
controlled \cite{tehranipoor2017dram}.

For RO-PUFs, a simple measurement-time approximation is

\begin{equation}
T_{\mathrm{RO}}
\approx
T_{\mathrm{window}}
+
T_{\mathrm{counter}}
+
T_{\mathrm{compare}},
\label{eq:ro_latency}
\end{equation}

where increasing $T_{\mathrm{window}}$ can improve frequency-estimation
confidence but directly increases authentication latency.

\subsection{Security Challenges and Limitations}
\label{sec:puf_challenges}

The principal limitation of a PUF is that physical uniqueness does not
automatically imply cryptographic unpredictability. A PUF may exhibit
excellent inter-device Hamming distance and near-perfect reliability
while remaining vulnerable to an algorithm that predicts unseen
responses. This distinction is especially important for strong PUFs
\cite{ruhrmair2010modeling,wisiol2022neural}.

Environmental sensitivity is a second major limitation. Temperature,
supply voltage, aging, device stress, and measurement conditions alter
physical characteristics and can increase false rejection unless the
system includes reliability screening, repeated sampling, or error
correction. Aging-resistant RO-PUF designs illustrate that long-term
stability may require circuit-level compensation rather than only
statistical post-processing \cite{rahman2015aging}.

Helper data introduces another attack surface. A design must state
whether helper information is public~\cite{becker2017helper}, authenticated, integrity
protected, or modifiable by the adversary. Practical fuzzy-extractor
work has shown that helper-data manipulation can invalidate security
assumptions that hold only for passive attackers
\cite{becker2017helper,pour2022helper}.

Side-channel leakage must also be considered. Response-dependent power,
timing, electromagnetic, or optical behavior can expose internal PUF
information even when external CRP access is restricted~\cite{mahmoud2013combined}. Physical
probing and fault injection can similarly target measurement logic,
challenge routing, response registers, helper-data processing, or ECC
decoding.

A complete PUF threat model should therefore specify whether the
adversary can

\begin{equation}
\mathcal{A}
=
\{
\mathcal{Q}_{C},
\mathcal{O}_{W},
\mathcal{M}_{E},
\mathcal{S}_{\mathrm{phys}},
\mathcal{F}_{\mathrm{inj}}
\},
\label{eq:puf_threat_capabilities}
\end{equation}

where $\mathcal{Q}_{C}$ denotes challenge-query capability,
$\mathcal{O}_{W}$ helper-data observation,
$\mathcal{M}_{E}$ environmental manipulation,
$\mathcal{S}_{\mathrm{phys}}$ side-channel observation, and
$\mathcal{F}_{\mathrm{inj}}$ fault-injection capability.

The security objective should then be evaluated under the strongest
adversary consistent with the intended application, rather than by
reporting uniqueness and reliability alone.

\FloatBarrier

\section{Hardware Obfuscation as a Functional Root of Trust}
\label{sec:locking}

\subsection{Complete Hardware-Obfuscation Processing Chain}
\label{sec:locking_pipeline}

Hardware obfuscation establishes a \emph{functional root of trust} by
making correct circuit behavior conditional on possession of an
authorized configuration or activation key. Early key-based hardware
activation schemes such as EPIC established the basic idea of
manufacturing a circuit whose intended functionality is enabled only
after secure post-fabrication activation \cite{roy2008epic,tehranipoor2019deep, mellor2021attacks, yue2021novel}. Unlike biometrics, which
establish trust in the user, and PUFs, which establish trust in a
physical device, hardware obfuscation controls whether the hardware
itself can execute its intended functionality. This distinction is
important in systems where possession of the physical implementation
should not automatically imply authorization to use, clone, reverse
engineer, or redistribute its protected functionality.

\begin{figure*}[t]
    \centering
    \includegraphics[width=0.98\textwidth]
    {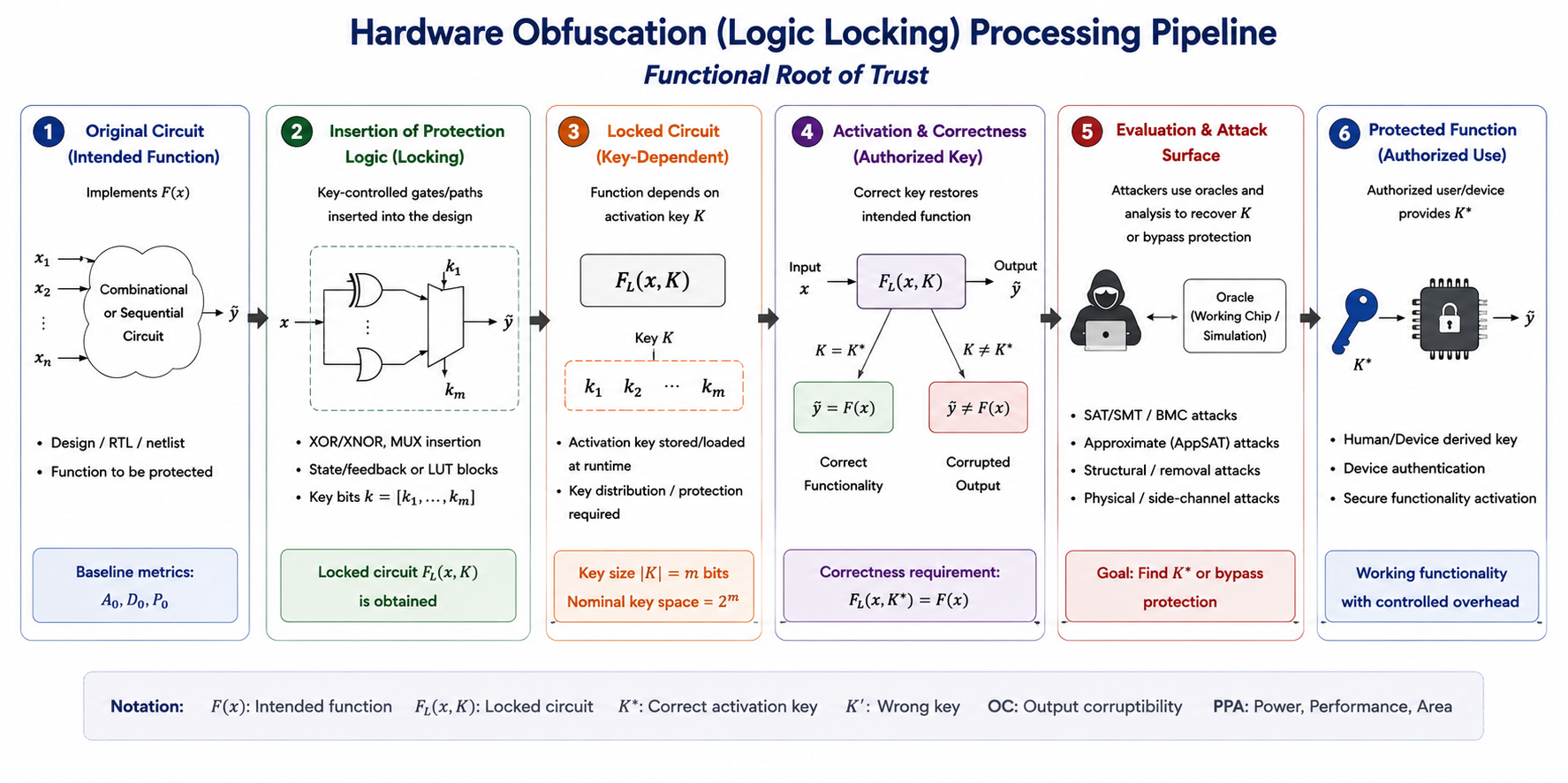}
    \caption{
    \textbf{Hardware-obfuscation and logic-locking processing pipeline as a functional root of trust.}
    The intended circuit $F(x)$ is first transformed by inserting key-controlled protection logic, such as XOR/XNOR gates, multiplexers, state-dependent elements, or programmable structures, to obtain the locked circuit $F_L(x,K)$. Correct functionality is restored only when the authorized activation key $K^{*}$ is supplied, whereas an incorrect key $K' \neq K^{*}$ should cause meaningful output corruption. The activation key may be securely stored, provisioned, or dynamically derived from trusted human- and device-specific information. The locked design must then be evaluated against SAT/SMT or oracle-guided attacks, approximate attacks, structural and removal attacks, and physical or side-channel key extraction. Successful authorization therefore requires both correct key activation and preservation of the intended functionality under acceptable implementation overhead.
    }
    \label{fig:hardware_obfuscation_pipeline}
\end{figure*}

Figure~\ref{fig:hardware_obfuscation_pipeline} illustrates the complete
logic-locking processing chain. An original circuit implementing the
intended functionality is first transformed through the insertion of
key-controlled gates, multiplexers, programmable structures, state
elements, or reconfigurable logic. The resulting circuit exhibits
key-dependent behavior and produces the correct output only when the
authorized activation key is supplied. The security of this
transformation must then be evaluated against oracle-guided,
approximate, structural, removal, and physical attacks while
simultaneously accounting for area, delay, power, and activation
overhead.

The overall transformation can be represented as

\begin{equation}
F(x)
\xrightarrow{\;\mathcal{L}(\cdot,K)\;}
F_L(x,K),
\label{eq:locking_transformation}
\end{equation}

where $\mathcal{L}(\cdot)$ denotes the locking or obfuscation
transformation and $K$ is an $m$-bit activation key, $K = [k_1,k_2,\ldots,k_m] \in \{0,1\}^{m}$. The central correctness requirement is

\begin{equation}
F_L(x,K^{*})
=
F(x),
\qquad
\forall x\in\mathcal{X},
\label{eq:lockcorrect}
\end{equation}

where $K^{*}$ denotes the authorized key and $\mathcal{X}$ is the
relevant functional input space.

For an incorrect key $K'$, a secure construction should instead
produce $F_L(x,K') \neq F(x)$ for a sufficiently large or security-critical portion of
$\mathcal{X}$. Thus, merely increasing the nominal key length does not
guarantee security. The effectiveness of the protection depends on how
the key influences circuit behavior, whether the protection structure
can be identified or removed, and how much information an attacker can
obtain from a working implementation.

\subsubsection{Intended Circuit and Protection Target}
\label{sec:locking_original}

The first stage in Fig.~\ref{fig:hardware_obfuscation_pipeline} is the
original hardware design. Let the intended combinational or sequential
circuit implement $y = F(x)$, where $x = [x_1,x_2,\ldots,x_n]$ is the functional input vector and $y$ denotes the intended output.

Before locking, the designer must identify what is being protected.
The target may be the entire circuit, a cryptographic accelerator, a
proprietary signal-processing block, a neural-network accelerator, an
authentication engine, or another security-critical function. The
choice of protection target directly affects both security and
implementation cost because inserting protection uniformly across the
design can introduce unnecessary overhead, whereas protecting only a
small identifiable region may facilitate structural removal attacks.

The unprotected design provides the baseline implementation metrics $\mathcal{M}_0 = \{A_0,D_0,P_0\}$, where $A_0$, $D_0$, and $P_0$ denote area, critical-path delay, and
power consumption, respectively. These values form the reference
against which the cost of hardware obfuscation is evaluated.

\subsubsection{Insertion of Key-Controlled Protection Logic}
\label{sec:locking_insertion}

The second stage modifies the hardware structure so that one or more
internal signals depend on the activation key. A general locking
transformation can be written as $F_L = \mathcal{L} \left( F,K,\Lambda \right)$, where $\Lambda$ describes the locations and types of inserted
protection structures.

For conventional XOR/XNOR locking, an internal signal $s_i$ can be
replaced by $\tilde{s}_i = s_i \oplus k_i$ or, depending on the intended correct key value, $\tilde{s}_i = s_i \odot k_i$, where $\oplus$ and $\odot$ denote XOR and XNOR, respectively.

MUX-based locking selects between alternative signal paths,

\begin{equation}
\tilde{s}_i
=
\begin{cases}
s_i^{(0)}, & k_i=0,\\
s_i^{(1)}, & k_i=1.
\end{cases}
\label{eq:mux_locking}
\end{equation}

Similar principles can be extended to state transitions, feedback
paths, LUT contents, programmable routing, or FPGA configuration bits.

If $m$ key-controlled elements are inserted, the nominal key space is $|\mathcal{K}| = 2^{m}$. However, this expression should not be interpreted as an attack
complexity of $2^m$. The SAT attack demonstrated that oracle-guided
analysis can often recover the correct key without exhaustive search
by iteratively eliminating large subsets of incorrect keys
\cite{subramanyan2015sat}. Hence, structural dependence between key
bits and circuit behavior is more important than nominal key-space
size alone.

\subsubsection{Locked Circuit and Key-Dependent Behavior}
\label{sec:locked_circuit}

After protection logic is inserted, the circuit implements the
key-dependent function $y_L = F_L(x,K)$. For the authorized key, $F_L(x,K^{*}) = F(x)$, whereas an incorrect key should ideally satisfy $F_L(x,K') \neq F(x), \qquad K'\neq K^{*}$, over a substantial portion of the input space.

A useful way to express the locking objective is therefore

\begin{equation}
\begin{aligned}
K=K^{*}
&\Rightarrow
F_L(x,K)=F(x),\\
K\neq K^{*}
&\Rightarrow
F_L(x,K)\neq F(x)
\quad\text{with high probability}.
\end{aligned}
\label{eq:locking_objective}
\end{equation}

This second condition is essential. Some SAT-resistant constructions
achieve high solver complexity while corrupting only a very small
fraction of input patterns. Such behavior may prevent exact key
recovery but still permit an adversary to obtain an approximately
correct implementation. Stripped-functionality locking, including
SFLL-HD, is representative of this design direction and illustrates why
solver resistance must be evaluated together with protected input
patterns, restore logic, and wrong-key corruptibility
\cite{hashemi2025srll,juretus2020increased}.

\subsubsection{Runtime Key Activation}
\label{sec:locking_activation}

The locked circuit becomes operational only after an activation key is
loaded or reconstructed. Let the runtime activation key be
$\hat{K}$. The activated output is $\hat{y} = F_L(x,\hat{K})$. Successful activation requires $\hat{K} = K^{*}$. In conventional implementations, the key may be stored in secure
nonvolatile memory, loaded during secure boot, or supplied through a
trusted provisioning mechanism. In the human--device--function
architecture considered in this survey, the activation key may instead
be reconstructed dynamically from biometric and PUF-derived
information,

\begin{equation}
K^{*}
=
\operatorname{KDF}
\left(
B_u
\parallel
R_d
\parallel
N
\parallel
ID_{\mathrm{app}}
\right),
\label{eq:bio_puf_lock_key}
\end{equation}

where $B_u$ denotes the biometric-derived credential, $R_d$ the
device-specific PUF response, $N$ a freshness value, and
$ID_{\mathrm{app}}$ an application identifier.

This construction changes the role of logic locking. The correct key
is no longer only a design secret; it becomes a runtime authorization
credential jointly tied to the human and device roots of trust.

The activation path itself therefore becomes a critical attack
surface. Even a mathematically strong locking primitive provides little
protection if an attacker can directly observe the key while it is
stored, transferred, loaded, or distributed through the design.
Runtime key registers, configuration buses, scan paths, debug
interfaces, and power or electromagnetic leakage must consequently be
included in the threat model.

\subsection{Logic-Locking Families}
\label{sec:locking_families}

Hardware-obfuscation techniques differ primarily in how they create
key dependence and in the adversarial assumption they seek to defeat.

\subsection{Oracle-Guided SAT Attacks}
\label{sec:sat_attack}

The SAT attack fundamentally changed the evaluation of logic locking
because it showed that nominal key-space size is not an adequate
security measure. The attacker is assumed to possess the locked
netlist $F_L(x,K)$ and access to a working implementation of the circuit that acts as an
oracle for the correct functionality $F(x)$
\cite{subramanyan2015sat}.

Two candidate keys $K_1$ and $K_2$ are distinguishable whenever there
exists an input $x_d$ such that $F_L(x_d,K_1) \neq F_L(x_d,K_2)$. Such an $x_d$ is called a distinguishing input pattern (DIP). The
oracle supplies $y_d = F(x_d)$, allowing all keys inconsistent with $(x_d,y_d)$ to be removed.

After $t$ oracle queries, the surviving key set can be represented as

\begin{equation}
\mathcal{K}_t
=
\left\{
K:
F_L(x_i,K)=F(x_i),
\;
i=1,\ldots,t
\right\}.
\label{eq:sat_key_set}
\end{equation}

The attack terminates when no additional distinguishing input can be
found and the remaining key candidates are functionally equivalent.

Consequently, SAT resistance should be reported using quantities such
as $T_{\mathrm{SAT}}, \qquad N_{\mathrm{DIP}}, \qquad N_{\mathrm{query}}$, rather than key length alone.

\subsection{Approximate Attacks}
\label{sec:approximate_attacks}

Exact reconstruction of $K^{*}$ is not always necessary for an
attacker. If an incorrect key reproduces the correct functionality for
most inputs, the protected design may still be commercially or
operationally useful.

For candidate key $K'$, approximate functional accuracy over input set
$\mathcal{X}$ can be defined as

\begin{equation}
\mathrm{FA}(K')
=
\frac{1}{|\mathcal{X}|}
\sum_{x\in\mathcal{X}}
\mathbb{1}
\left[
F_L(x,K')
=
F(x)
\right].
\label{eq:functional_accuracy}
\end{equation}

Correspondingly, $\mathrm{OC}(K') = 1-\mathrm{FA}(K')$ is the wrong-key output corruptibility.

AppSAT demonstrated that an adversary can intentionally terminate
before exact SAT convergence and recover a key that provides high
approximate functional correctness
\cite{shamsi2017appsat, mellor2021attacks}. This is particularly relevant for locking
schemes whose SAT hardness is obtained by making incorrect keys differ
from the correct implementation only for rare input patterns.

The security requirement is therefore not merely $K'\neq K^{*}$, but rather that incorrect keys exhibit sufficiently low useful
functionality.

\subsection{Structural and Removal Attacks}
\label{sec:removal_attacks}

Oracle resistance alone is also insufficient if the protection
circuitry can be identified directly from the netlist. Structural
attacks exploit topology, gate types, fan-in/fan-out behavior,
symmetry, reconvergent paths, or synthesis artifacts to distinguish
locking logic from the original functional design.

Let $\mathcal{L}_{P} \subseteq \mathcal{G}_{L}$ denote the set of gates implementing protection within locked netlist
$\mathcal{G}_{L}$. A removal attack seeks an estimate $\hat{\mathcal{L}}_{P} \approx \mathcal{L}_{P}$. A useful removal-success metric can therefore be defined as

Let $\mathcal{E}_{\mathrm{rem}}$ denote the event that an attacker
restores useful functionality after removing or bypassing the
protection logic. The corresponding removal-success probability is
\begin{equation}
P_{\mathrm{rem}}
=
\Pr\!\left[\mathcal{E}_{\mathrm{rem}}\right].
\label{eq:removal_success}
\end{equation}

Recent attacks on SAT-hard constructions reinforce the distinction
between resistance to one solver strategy and resistance to general
structural analysis \cite{removal2024,srll2025}. Consequently,
``SAT-resistant'' should not be used as a synonym for
``secure logic locking.''

\subsection{Physical and Side-Channel Attacks}
\label{sec:locking_physical}

A further attack class targets the activation mechanism rather than the
locking algorithm. The correct key must eventually exist physically in
registers, memory, routing resources, configuration logic, or
cryptographic state~\cite{picek2023sok,jeon2021new, tehranipoor2023breaking}.

An attacker may therefore attempt to recover $K^{*}$ through power analysis, electromagnetic analysis, timing leakage,
probing, scan-chain access, debug interfaces, fault injection, or
configuration-memory extraction. The effective security of a locked circuit can be viewed conceptually
as

\begin{equation}
S_{\mathrm{effective}}
=
\min
\left\{
S_{\mathrm{logical}},
S_{\mathrm{structural}},
S_{\mathrm{physical}},
S_{\mathrm{keypath}}
\right\},
\label{eq:effective_lock_security}
\end{equation}

because compromising any one of these components may be sufficient to
recover or bypass the protected functionality.

This issue becomes especially important when biometric or PUF-derived
keys are used for activation. The biometric and PUF primitives may be
secure individually, yet their benefit is lost if the derived
$K_{u,d}$ is exposed between the key-derivation unit and the locked
hardware~\cite{galbally2020new}.

\subsection{Obfuscation Evaluation Metrics}
\label{sec:locking_metrics}

Hardware-obfuscation evaluation should jointly measure functional
corruption, attack resistance, and implementation overhead.

\subsubsection{Wrong-Key Output Corruptibility}

For a wrong key $K'$ and an input set $\mathcal{X}$, output
corruptibility is

\begin{equation}
\mathrm{OC}(K')
=
\frac{1}{|\mathcal{X}|}
\sum_{x\in\mathcal{X}}
\mathbb{1}
\left[
F_L(x,K')
\neq
F(x)
\right].
\label{eq:corruption}
\end{equation}

A population-level quantity over wrong keys can be written as

\begin{equation}
\overline{\mathrm{OC}}
=
\frac{1}{
|\mathcal{K}|-1
}
\sum_{
K'\in
\mathcal{K}\setminus\{K^{*}\}
}
\mathrm{OC}(K').
\label{eq:mean_corruption}
\end{equation}

High average corruption is desirable, although the minimum or
lower-tail corruption can also be important because an attacker only
needs to find one approximately useful key.

Thus, $\mathrm{OC}_{\min} = \min_{ K'\neq K^{*} } \mathrm{OC}(K')$ can expose weak keys hidden by a favorable average.

\subsubsection{Key Sensitivity}

Key sensitivity measures how strongly changing activation bits alters
the circuit output. For two keys $K_1$ and $K_2$, one possible
definition is

\begin{equation}
\mathrm{KS}(K_1,K_2)
=
\frac{1}{|\mathcal{X}|}
\sum_{x\in\mathcal{X}}
\mathbb{1}
\left[
F_L(x,K_1)
\neq
F_L(x,K_2)
\right].
\label{eq:key_sensitivity}
\end{equation}

To examine sensitivity around the correct key, define
$K^{*(i)}$ as $K^{*}$ with bit $i$ inverted. The average single-bit
key sensitivity is then

\begin{equation}
\mathrm{KS}_{1}
=
\frac{1}{m}
\sum_{i=1}^{m}
\mathrm{OC}
\left(
K^{*(i)}
\right).
\label{eq:single_bit_key_sensitivity}
\end{equation}

A low value indicates that some key bits have little functional
influence and may therefore be structurally or functionally weak.

\subsubsection{Attack Success and Complexity}

For attack algorithm $\mathcal{A}$, exact key-recovery success can be
written as

\begin{equation}
P_{\mathrm{KR}}^{\mathcal{A}}
=
\Pr
\left[
\hat{K}_{\mathcal{A}}
=
K^{*}
\right].
\label{eq:attack_key_recovery}
\end{equation}

Because approximate functionality may also constitute a successful
attack, one may define

\begin{equation}
P_{\mathrm{approx}}^{\mathcal{A}}(\tau)
=
\Pr
\left[
\mathrm{FA}
\left(
\hat{K}_{\mathcal{A}}
\right)
\geq
\tau
\right],
\label{eq:approx_attack_success}
\end{equation}

where $\tau$ specifies the minimum functionality useful to the
attacker.

Evaluation should therefore report at least

\begin{equation}
\left\{
T_{\mathrm{attack}},
N_{\mathrm{query}},
N_{\mathrm{DIP}},
P_{\mathrm{KR}},
P_{\mathrm{approx}}
\right\}.
\label{eq:attack_metric_set}
\end{equation}

Timeouts must also be reported explicitly because an attack that fails
within one arbitrary experimental limit is not equivalent to a
provable security bound.

\subsubsection{Power--Performance--Area Overhead}

Hardware protection competes directly with conventional implementation
objectives. Let the unlocked design have area $A_0$, critical-path
delay $D_0$, and power $P_0$, while the locked implementation has $A_L$, $D_L$, and $P_L$. Normalized overheads are $\Delta A = \frac{A_L-A_0}{A_0}$, $\Delta D = \frac{D_L-D_0}{D_0}$, and $\Delta P = \frac{P_L-P_0}{P_0}$. These quantities may be expressed as fractions or percentages.

The total runtime cost should additionally include key loading or
reconstruction,

\begin{equation}
T_{\mathrm{activation}}
=
T_{\mathrm{key}}
+
T_{\mathrm{load}}
+
T_{\mathrm{config}}
+
T_{\mathrm{verify}},
\label{eq:activation_latency}
\end{equation}

particularly for FPGA/eFPGA and bitstream-based protection where
reconfiguration time may be non-negligible.

For the integrated architecture,

\begin{equation}
T_{\mathrm{key}}
=
T_{\mathrm{bio}}
+
T_{\mathrm{PUF}}
+
T_{\mathrm{KDF}}
+
T_{\mathrm{ECC}},
\label{eq:joint_activation_latency}
\end{equation}

showing that the cost of functional authorization cannot always be
separated from the human and device roots of trust.

\subsection{Security--Corruption--Cost Trade-Off}
\label{sec:locking_tradeoff}

Hardware obfuscation is fundamentally a multi-objective design
problem. A useful protection mechanism should simultaneously increase
attack complexity, maintain high corruption for incorrect keys, and
minimize implementation overhead.

Conceptually, the design objective can be expressed as

\begin{equation}
\begin{aligned}
\mathcal{L}^{*}
=
\arg\max_{\mathcal{L}}
\big[
&
\lambda_1 S_{\mathrm{attack}}
+
\lambda_2 \overline{\mathrm{OC}}
+
\lambda_3 \mathrm{KS}
\\
&
-
\lambda_4 \Delta A
-
\lambda_5 \Delta D
-
\lambda_6 \Delta P
\big],
\end{aligned}
\label{eq:locking_multiobjective}
\end{equation}

where $S_{\mathrm{attack}}$ represents resistance to the considered
attack model and the coefficients $\lambda_i$ capture
application-dependent priorities.

The competing objectives explain why no single logic-locking family is
universally preferable. Random XOR/XNOR locking can provide low
implementation overhead and substantial corruption but is vulnerable
to SAT attacks. SARLock and Anti-SAT-like constructions increase
solver complexity but may create low-corruption wrong keys or
identifiable protection structures
\cite{yasin2016sarlock,xie2019antisat}. Approximation-aware designs
attempt to improve this balance but generally increase implementation
complexity. Reconfigurable approaches can conceal larger regions of
logic but impose greater area, routing, configuration-memory, and
bitstream-security requirements.

\subsection{Challenges and Limitations}
\label{sec:locking_challenges}

Logic locking cannot be assessed reliably through a single metric because its security depends on the interaction among key-space size, functional corruption, structural properties, solver resistance, implementation transformations, and runtime key protection. Although an $m$-bit key nominally defines a search space of $2^m$ candidates, key length alone is not a meaningful measure of practical security. Structural analysis, oracle-guided reasoning, and solver-based attacks can often eliminate large portions of the nominal key space without exhaustively enumerating candidate keys. As a result, a longer key does not necessarily imply a proportionally stronger locking scheme.

A similar limitation applies to SAT resistance. Forcing a SAT solver to perform many iterations does not necessarily guarantee that the protected circuit remains functionally secure. Some schemes can substantially increase exact key-recovery difficulty while still allowing incorrect keys to reproduce most of the intended functionality. Approximate attacks exploit this weakness by seeking keys that are not exact but nevertheless achieve high functional agreement with the original circuit \cite{shamsi2017appsat}. Consequently, solver runtime and timeout behavior should be interpreted together with approximate functional correctness rather than treated as sufficient evidence of security.

Wrong-key output corruption is also incomplete as a standalone metric. A design may produce substantial corruption under random incorrect keys while still containing a structurally identifiable protection cone, restore unit, or key-dependent subcircuit that can be removed, isolated, or bypassed. Functional distortion therefore needs to be considered together with structural and removal resistance. A locking mechanism is not robust if its protection logic can be recognized and neutralized without recovering the intended key.

The security of a locking construction can also change after synthesis and physical implementation. Logic optimization may simplify inserted gates, alter key-gate observability, collapse redundant structures, introduce recognizable signatures, or otherwise modify the intended protection topology. Accordingly, evaluating only the RTL-level construction can overestimate security. A meaningful assessment should examine at least the synthesized netlist, and ideally the post-layout implementation when physical structure, routing, timing, or side-channel behavior is part of the threat model.

Runtime key protection remains equally important because logic locking does not eliminate the secret; it changes the way the secret controls functionality. Even a theoretically strong locking scheme can be bypassed if the authorized key $K^{*}$ is exposed through nonvolatile memory, scan chains, debug interfaces, configuration buses, physical probing, fault injection, or timing, power, and electromagnetic side channels. In such cases, the attacker does not need to solve the locking problem at all. The effective functional root of trust therefore depends not only on the obfuscation algorithm but also on the complete key-generation, delivery, storage, activation, and erasure path.

These limitations reveal a broader design tension among attack resistance, wrong-key functional corruption, and implementation overhead. Improving one property can weaken another: mechanisms designed to increase solver complexity may reduce output corruption, while stronger corruption or structural protection may require additional logic, routing, control, or activation circuitry. These limitations reveal a broader design tension among attack
resistance, wrong-key functional corruption, and implementation
overhead. Improving one property can weaken another: mechanisms that
increase solver resistance may reduce output corruption, whereas
stronger corruption or structural protection may require additional
logic, routing, or activation circuitry. Therefore, these properties
should be evaluated jointly rather than interpreted in isolation.

For this reason, a rigorous comparison of hardware-obfuscation schemes should consider multiple dimensions jointly. Relevant evidence includes exact and approximate attack success, SAT/SMT/BMC runtime and timeout policy, oracle-query complexity, the number of distinguishing inputs, functional correctness under incorrect keys, wrong-key output corruption, key sensitivity, structural and removal-attack success, and the security of the activation path. When implementation efficiency is also within scope, activation latency and power--performance--area overhead can be reported separately rather than conflated with security. Only such a combined evaluation can determine whether a locking scheme provides a practical functional root of trust rather than merely resisting a particular attack algorithm.
\section{Composed Trust Architectures}
\label{sec:composition}

The preceding sections treat biometrics, PUFs, and hardware
obfuscation as three distinct roots of trust. Their composition creates
stronger security properties because authorization can be conditioned
on more than one independent source: a biometric identifies the
\emph{human}, a PUF identifies the \emph{physical device}, and
hardware obfuscation controls the \emph{function} that can execute.
The resulting architectures range from two-root bindings to a complete
human--device--function trust chain.

\begin{figure*}[t]
    \centering
    \includegraphics[width=0.98\textwidth]
    {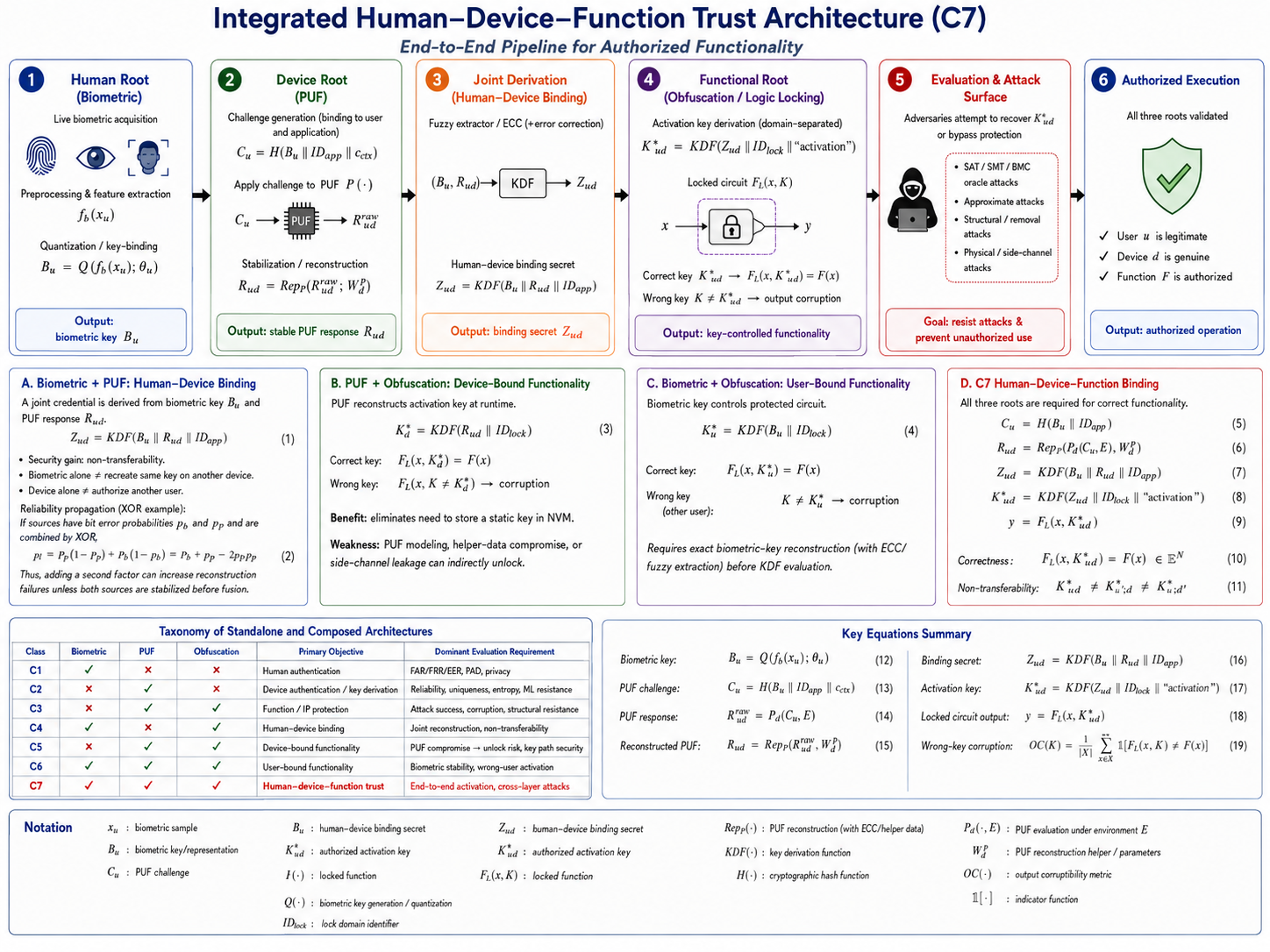}
    \caption{
    \textbf{Integrated human--device--function trust architecture.}
    The biometric layer reconstructs a stable user-dependent representation $B_u$; the PUF layer generates and stabilizes a device-dependent response $R_{u,d}$; and a cryptographic derivation stage fuses both values into the binding secret $Z_{u,d}$. A domain-separated activation key $K_{u,d}^{*}$ then controls the locked function $F_L(x,K)$. Correct execution therefore depends on successful reconstruction at both upstream roots and on secure delivery of the derived activation key. The figure also highlights pairwise compositions and the principal cross-layer interfaces that must be protected against replay, modeling, helper-data manipulation, key leakage, and locking bypass.
    }
    \label{fig:integration_pipeline}
\end{figure*}

Figure~\ref{fig:integration_pipeline} illustrates the complete
integration pipeline. A live biometric measurement is first converted
into a reproducible biometric representation. This representation can
be used to derive a PUF challenge or can be combined with a stabilized
PUF response. A cryptographic key-derivation function then creates a
joint human--device binding secret. Finally, this binding secret is
used to derive or recover the hardware activation key required by the
protected circuit.

\subsection{Biometric + PUF: Human--Device Binding}
\label{sec:bio_puf_binding}

The first composition couples a biometric-derived representation with
a device-specific PUF response. The purpose is to prevent either
credential from being independently transferable. Possession of a
biometric representation alone should not reproduce the same
credential on another device, while possession of the authentic
device alone should not authorize a different user.

Let the live biometric measurement from user $u$ be $\mathbf{x}_u$, and let the biometric subsystem produce $B_u = Q \left( f_{\theta}(\mathbf{x}_u); \Theta_u \right)$, where $f_{\theta}(\cdot)$ denotes preprocessing and feature extraction,
$Q(\cdot)$ denotes the biometric key-generation or binding operation,
and $\Theta_u$ represents enrolled reliability information, selected
features, quantization parameters, or helper data.

Because a PUF accepts a challenge, the biometric representation can
participate directly in challenge generation. A domain-separated
challenge can be constructed as

\begin{equation}
C_u
=
H
\left(
B_u
\parallel
ID_{\mathrm{app}}
\parallel
c_{\mathrm{ctx}}
\right),
\label{eq:composition_puf_challenge}
\end{equation}

where $ID_{\mathrm{app}}$ identifies the protected application and
$c_{\mathrm{ctx}}$ denotes optional non-secret contextual information.

Device $d$ then evaluates $R_{u,d}^{\mathrm{raw}} = P_d \left( C_u,E \right)$, where $E$ represents environmental conditions.

Because the PUF response is noisy, the raw value should not normally
be passed directly into a cryptographic derivation function. Instead,
a reconstruction operation produces $R_{u,d} = \operatorname{Rep}_{P} \left( R_{u,d}^{\mathrm{raw}}, W_d \right)$, where $W_d$ denotes PUF helper information.

A stable human--device binding secret can then be generated as

\begin{equation}
\boxed{
Z_{u,d}
=
\operatorname{KDF}
\left(
B_u
\parallel
R_{u,d}
\parallel
ID_{\mathrm{app}}
\right)
}
\label{eq:human_device_binding_key}
\end{equation}

rather than simply concatenating or XORing the two sources.

The desired binding property is $Z_{u_1,d} \neq Z_{u_2,d}, \qquad u_1\neq u_2$, and $Z_{u,d_1} \neq Z_{u,d_2}, \qquad d_1\neq d_2$. Thus, the credential is simultaneously dependent on the identity of
the user and on the physical characteristics of the device.

\subsubsection{Reliability Propagation}
\label{sec:composition_reliability}

Composition does not automatically improve reliability. Suppose the
biometric subsystem has bit-error probability $p_B$ and the PUF has
bit-error probability $p_P$. If the two raw binary sources are
combined directly using XOR, then the probability that a resulting bit
is incorrect is

\begin{equation}
\begin{aligned}
p_J
&=
p_B(1-p_P)
+
p_P(1-p_B)
\\
&=
p_B+p_P-2p_Bp_P.
\end{aligned}
\label{eq:jointber}
\end{equation}

This equation is useful for illustrating error propagation in direct
bitwise fusion. It should \emph{not}, however, be interpreted as the
error probability of a cryptographic KDF. Cryptographic hash and KDF
functions exhibit avalanche behavior, such that even a single
incorrect input bit may produce an unrelated output.

Consequently, both sources should first be reconstructed into their
stable forms, $\hat{B}_u=B_u$ and $\hat{R}_{u,d}=R_{u,d}$, before cryptographic fusion.

Let $P_B = \Pr[\hat{B}_u=B_u]$ denote successful biometric-key reconstruction and $P_P = \Pr[\hat{R}_{u,d}=R_{u,d}]$ denote successful PUF reconstruction.

Under a simplified independence assumption, successful reconstruction
of the binding secret satisfies approximately $P_{\mathrm{bind}} \approx P_B P_P$. More generally,

\begin{equation}
P_{\mathrm{bind}}
=
\Pr
\left[
\hat{B}_u=B_u
\;\land\;
\hat{R}_{u,d}=R_{u,d}
\right],
\label{eq:binding_success_general}
\end{equation}

which should be measured directly when the two error processes can be
correlated.

\subsubsection{Implementation}

A practical human--device binding implementation separates
\emph{enrollment} from \emph{runtime reconstruction}. During
enrollment, multiple biometric observations are collected to establish
the user's stable quantization parameters $\Theta_u$ and, where
required, biometric helper information $W_u^{B}$. The PUF is also
characterized under the expected operating range to identify reliable
response positions and construct PUF helper information $W_d^{P}$.

The enrollment state can therefore be represented abstractly as

\begin{equation}
\mathcal{S}_{u,d}
=
\left\{
\Theta_u,
W_u^{B},
W_d^{P},
ID_{\mathrm{app}}
\right\}.
\label{eq:binding_enrollment_state}
\end{equation}

Importantly, neither the raw biometric measurement nor the raw
device-derived secret needs to be stored as plaintext.

At runtime, the implementation performs

\begin{equation}
\mathbf{x}_u
\rightarrow
\hat{B}_u
\rightarrow
C_u
\rightarrow
R_{u,d}^{\mathrm{raw}}
\rightarrow
\hat{R}_{u,d}
\rightarrow
\hat{Z}_{u,d}.
\label{eq:bio_puf_runtime}
\end{equation}

Only after successful reconstruction of both factors is the final
binding secret available.

\subsection{PUF + Hardware Obfuscation:
Device-Bound Functionality}
\label{sec:puf_obfuscation}

The second composed architecture makes correct hardware functionality
dependent on the physical device. Instead of storing a conventional
logic-locking key directly in nonvolatile memory, the key is
reconstructed from a PUF response.

Let the enrolled PUF-derived secret be $R_d = \operatorname{Rep}_{P} \left( P_d(C,E), W_d^{P} \right)$. A device-specific activation key is derived as

\begin{equation}
\boxed{
K_d^{*}
=
\operatorname{KDF}
\left(
R_d
\parallel
ID_{\mathrm{lock}}
\right)
}
\label{eq:puf_lock_key}
\end{equation}

where $ID_{\mathrm{lock}}$ provides domain separation for the protected
hardware block.

The locked circuit executes $y = F_L \left( x,K_d^{*} \right)$. Correctness requires $F_L \left( x,K_d^{*} \right) = F(x), \qquad \forall x\in\mathcal{X}$. If the same netlist is transferred to another physical device
$d'\neq d$, the expected PUF response changes, $R_{d'} \neq R_d$, and consequently $K_{d'}^{*} \neq K_d^{*}$ with high probability.

The copied design therefore remains locked even when the attacker
possesses the complete protected netlist.

\subsubsection{Implementation}

During trusted provisioning, the designer determines the activation
key associated with the enrolled PUF response and configures the
locking structure accordingly. Runtime operation reconstructs the PUF
response, derives the key, loads it into the locking network, and
clears transient key material after activation.

The hardware path can be represented as

\begin{equation}
\boxed{
P_d
\rightarrow
\operatorname{Rep}_{P}
\rightarrow
R_d
\rightarrow
\operatorname{KDF}
\rightarrow
K_d^{*}
\rightarrow
F_L
}
\label{eq:puf_obf_pipeline}
\end{equation}

The interface between the KDF and the locking network must remain
inside the trusted hardware boundary. Otherwise, probing, scan access,
debug interfaces, or side-channel leakage can bypass the PUF entirely
by recovering $K_d^{*}$ after reconstruction.

A further requirement is integrity protection for helper information.
If an adversary can maliciously alter $W_d^{P}$, the reconstructed
response may be corrupted before the activation key is derived.

\subsection{Biometric + Hardware Obfuscation:
User-Bound Functionality}
\label{sec:bio_obfuscation}

A biometric credential can also control protected functionality
without a PUF. In this architecture, successful biometric
reconstruction directly determines whether the hardware is unlocked.

After reliable biometric reconstruction, $B_u = \operatorname{Rep}_{B} \left( \tilde{B}_u, W_u^{B} \right)$, the hardware activation key can be derived as $K_u^{*} = \operatorname{KDF} \left( B_u \parallel ID_{\mathrm{lock}} \right)$. The protected hardware then implements $y = F_L \left( x,K_u^{*} \right)$. For authorized user $u^{*}$, $F_L \left( x,K_{u^{*}}^{*} \right) = F(x)$, whereas a biometric representation from another user should derive a
different key, $u\neq u^{*} \Rightarrow K_u^{*} \neq K_{u^{*}}^{*}$. This architecture changes the conventional biometric decision model.
Instead of producing only $\mathrm{Accept}/\mathrm{Reject}$, the biometric measurement produces cryptographic material that is
required for functional correctness.

This stronger coupling also means that biometric reconstruction must
be exact before the KDF is evaluated. Approximate biometric similarity
is insufficient because

\begin{equation}
B_u'
\neq
B_u
\quad\Longrightarrow\quad
\operatorname{KDF}(B_u')
\not\approx
\operatorname{KDF}(B_u).
\end{equation}

Consequently, quantization, fuzzy extraction, secure sketches, or
other error-tolerant reconstruction must precede the hardware
activation stage.

\subsection{Biometric + PUF + Obfuscation:
C7 Human--Device--Function Binding}
\label{sec:c7}

The C7 architecture combines all three roots of trust. Its security goal is stronger than conventional multifactor authentication: correct functionality should emerge only when an authorized human interacts with the intended physical device and the resulting joint credential activates the protected function.

The complete architecture can be expressed in four stages.

\subsubsection{Stage 1: Human-Derived Secret}

A live biometric acquisition produces $B_u = Q \left( f_{\theta}(\mathbf{x}_u); \Theta_u \right)$. If fuzzy reconstruction is used, $B_u = \operatorname{Rep}_{B} \left( \tilde{B}_u, W_u^{B} \right)$.
\subsubsection{Stage 2: Biometric-Dependent Device Response}

The biometric-derived value can define the PUF challenge, $C_u = H \left( B_u \parallel ID_{\mathrm{app}} \right)$, and the physical device generates $R_{u,d}^{\mathrm{raw}} = P_d \left( C_u,E \right)$. After response stabilization, $R_{u,d} = \operatorname{Rep}_{P} \left( R_{u,d}^{\mathrm{raw}}, W_d^{P} \right)$.
\subsubsection{Stage 3: Human--Device Binding Secret}

The stable biometric and PUF components are fused cryptographically,

\begin{equation}
\boxed{
Z_{u,d}
=
\operatorname{KDF}
\left(
B_u
\parallel
R_{u,d}
\parallel
ID_{\mathrm{app}}
\right)
}
\label{eq:c7_binding_secret}
\end{equation}

where $Z_{u,d}$ is a persistent human--device binding secret.

The use of a stable binding secret is important. A conventional static
logic-locking circuit expects the same correct activation key at each
execution. Therefore, a changing session nonce should generally not be
inserted directly into the static locking key unless the hardware is
explicitly designed for dynamic reconfiguration.

\subsubsection{Stage 4: Function-Specific Activation Key}

A domain-separated activation key is derived from the stable binding
secret,

\begin{equation}
\boxed{
K_{u,d}^{*}
=
\operatorname{KDF}
\left(
Z_{u,d}
\parallel
ID_{\mathrm{lock}}
\parallel
\texttt{"activation"}
\right).
}
\label{eq:c7_activation_key}
\end{equation}

The locked hardware then computes $y = F_L \left( x,K_{u,d}^{*} \right)$. Correct operation requires $F_L \left( x,K_{u^{*},d^{*}}^{*} \right) = F(x), \qquad \forall x\in\mathcal{X}$. An unauthorized user should derive $K_{u,d^{*}}^{*} \neq K_{u^{*},d^{*}}^{*}, \qquad u\neq u^{*}$, while execution on another device should produce $K_{u^{*},d}^{*} \neq K_{u^{*},d^{*}}^{*}, \qquad d\neq d^{*}$. Thus,

\begin{equation}
\boxed{
\mathrm{CorrectFunction}
\Longrightarrow
(u,d,F)
=
(u^{*},d^{*},F^{*})
}
\label{eq:c7_binding_condition}
\end{equation}

under the security assumptions of the biometric, PUF, KDF, and
obfuscation mechanisms.

\subsubsection{Session Freshness Without Changing the Static Lock Key}
\label{sec:c7_freshness}

Freshness should be separated from the static activation key. Once the
stable human--device binding secret $Z_{u,d}$ is available, a
session-specific authenticator can be generated as

\begin{equation}
A_t
=
\operatorname{MAC}_{Z_{u,d}}
\left(
N_t
\parallel
ID_{\mathrm{app}}
\parallel
\mathrm{ctr}_t
\right),
\label{eq:c7_session_authenticator}
\end{equation}

where $N_t$ is a fresh nonce and $\mathrm{ctr}_t$ is an optional
monotonic counter.

Thus, $N_t\neq N_{t+1} \Rightarrow A_t\neq A_{t+1}$, while the hardware activation key remains $K_{u,d}^{*} = \mathrm{constant}$ for the same authorized human--device--function binding.

This separation simultaneously supports replay resistance and
compatibility with conventional static logic locking.

\subsubsection{Alternative: Wrapped Activation Key}

A second implementation avoids directly configuring the locking
network from biometric and PUF outputs. A random activation key
$K_L^{*}$ can first be chosen during trusted provisioning. The key is
then cryptographically wrapped under the binding secret, $C_L = \operatorname{Enc}_{Z_{u,d}} \left( K_L^{*} \right)$. At runtime, correct reconstruction of $Z_{u,d}$ enables $\hat{K}_L = \operatorname{Dec}_{Z_{u,d}} \left( C_L \right)$. The protected circuit operates correctly only if $\hat{K}_L = K_L^{*}$. This implementation is particularly useful when the logic-locking key
must remain fixed but the biometric and PUF mechanisms are used to
control access to it.

\subsubsection{Runtime Integration Protocol}

The complete C7 runtime sequence is therefore

\begin{equation}
\boxed{
\begin{aligned}
\mathbf{x}_u
&\rightarrow
B_u
\rightarrow
C_u
\rightarrow
R_{u,d}^{\mathrm{raw}}
\\
&\rightarrow
R_{u,d}
\rightarrow
Z_{u,d}
\rightarrow
K_{u,d}^{*}
\rightarrow
F_L.
\end{aligned}
}
\label{eq:c7_complete_pipeline}
\end{equation}

More explicitly,

\begin{align}
B_u
&=
\operatorname{Rep}_{B}
\left(
Q(f_{\theta}(\mathbf{x}_u)),
W_u^{B}
\right),
\label{eq:c7_full1}
\\
C_u
&=
H
\left(
B_u\parallel ID_{\mathrm{app}}
\right),
\label{eq:c7_full2}
\\
R_{u,d}
&=
\operatorname{Rep}_{P}
\left(
P_d(C_u,E),
W_d^{P}
\right),
\label{eq:c7_full3}
\\
Z_{u,d}
&=
\operatorname{KDF}
\left(
B_u\parallel R_{u,d}\parallel ID_{\mathrm{app}}
\right),
\label{eq:c7_full4}
\\
K_{u,d}^{*}
&=
\operatorname{KDF}
\left(
Z_{u,d}\parallel ID_{\mathrm{lock}}
\right),
\label{eq:c7_full5}
\\
y
&=
F_L
\left(
x,K_{u,d}^{*}
\right).
\label{eq:c7_full6}
\end{align}

\subsubsection{End-to-End Correctness}

Legitimate activation requires successful reconstruction at every
upstream stage. Define $\mathcal{E}_{B} = \{ \hat{B}_u=B_u \}$, $\mathcal{E}_{P} = \{ \hat{R}_{u,d}=R_{u,d} \}$, and $\mathcal{E}_{K} = \{ \hat{K}_{u,d}^{*}=K_{u,d}^{*} \}$. The end-to-end successful activation probability is

\begin{equation}
P_{\mathrm{act}}
=
\Pr
\left[
\mathcal{E}_{B}
\cap
\mathcal{E}_{P}
\cap
\mathcal{E}_{K}
\right].
\label{eq:c7_activation_probability}
\end{equation}

No independence assumption is necessary in this definition.

The legitimate activation failure rate is $\mathrm{LAFR} = 1-P_{\mathrm{act}}$. For unauthorized subjects or devices, define

\begin{equation}
\mathrm{UAR}
=
\Pr
\left[
F_L
\left(
x,\hat{K}_{a,d'}
\right)
=
F(x)
\;\middle|\;
(a,d')
\neq
(u^{*},d^{*})
\right],
\label{eq:c7_uar}
\end{equation}

where UAR denotes the \emph{unauthorized activation rate}.

A desirable integrated architecture simultaneously seeks $\mathrm{LAFR} \rightarrow 0$ and $\mathrm{UAR} \rightarrow 0$.
\subsubsection{Implementation Trust Boundaries}

The integrated architecture introduces interfaces that do not exist in
the individual primitives. These interfaces should be treated as
explicit trust boundaries.

The biometric-to-PUF boundary carries either $B_u$ or a value derived
from it. Compromise at this interface may permit replay or controlled
PUF querying. The PUF-to-KDF boundary carries the reconstructed device
response $R_{u,d}$ and must prevent CRP harvesting and response
exposure. The KDF-to-locking boundary carries the most sensitive
derived value because observation of $K_{u,d}^{*}$ can bypass both the
biometric and PUF roots of trust.

The principal runtime secrets should therefore have intentionally
short lifetimes. Conceptually, $B_u \xrightarrow{\mathrm{use}} \operatorname{erase}$, $R_{u,d} \xrightarrow{\mathrm{use}} \operatorname{erase}$, and $K_{u,d}^{*} \xrightarrow{\mathrm{activation}} \operatorname{erase}$ whenever the underlying hardware allows ephemeral reconstruction.

Intermediate values should not be exposed through software-visible
buffers, debug interfaces, scan chains, or externally accessible
configuration buses unless they are independently protected.

\subsection{Cross-Layer Failure Propagation}
\label{sec:composition_failure}

Composition also means that compromise of one root can propagate into
another. Let $\mathcal{A}_{B}$ represent successful biometric spoofing or biometric-key recovery, $\mathcal{A}_{P}$ successful prediction or reconstruction of the PUF response, and $\mathcal{A}_{L}$ successful recovery or bypass of the hardware activation mechanism.

An end-to-end attack chain is therefore better represented by

\begin{equation}
P_{\mathrm{chain}}
=
\Pr
\left[
\mathcal{A}_{B}
\cap
\mathcal{A}_{P}
\cap
\mathcal{A}_{L}
\right]
\label{eq:c7_attack_chain}
\end{equation}

than by multiplying isolated attack probabilities unless statistical
independence has been experimentally justified.

For example, leakage of $B_u$ can expose the challenge used to query
the PUF; successful PUF modeling can then reproduce $R_{u,d}$; and
recovery of both values may permit derivation of $Z_{u,d}$ and the
hardware activation key. Conversely, direct extraction of
$K_{u,d}^{*}$ from the activation path may bypass both upstream roots
entirely.

This motivates evaluation of the complete composition rather than
claiming

\begin{equation}
\mathrm{Secure}
=
\mathrm{Secure}_{B}
\land
\mathrm{Secure}_{P}
\land
\mathrm{Secure}_{L}.
\end{equation}

Instead, security must include the composition and its interfaces,

\begin{equation}
\boxed{
\mathrm{Secure}_{\mathrm{E2E}}
=
\mathrm{Secure}
\left(
B
\circ
P
\circ
\operatorname{KDF}
\circ
F_L
\right).
}
\label{eq:composition_security}
\end{equation}

\subsection{Taxonomy of Standalone and Composed Architectures}

Table~\ref{tab:classes} summarizes the seven architectural classes.
The classification distinguishes which roots of trust participate in
the authorization decision and, for composed systems, which
cross-layer property must be evaluated.

\begin{table*}[t]
\centering
\caption{Taxonomy of standalone and composed human--device--function
trust architectures.}
\label{tab:classes}
\small
\begin{tabularx}{0.94\textwidth}{
c c c c
p{0.25\textwidth}
X}
\toprule
\textbf{Class} &
\textbf{Bio.} &
\textbf{PUF} &
\textbf{Obf.} &
\textbf{Primary objective} &
\textbf{Dominant evaluation requirement}
\\
\midrule

C1 &
\cmark & \xmark & \xmark &
Human authentication &
FAR/FRR/EER, PAD, biometric privacy
\\

C2 &
\xmark & \cmark & \xmark &
Device authentication / key derivation &
Reliability, uniqueness, entropy, modeling resistance
\\

C3 &
\xmark & \xmark & \cmark &
Functional/IP protection &
Exact and approximate attack success, corruption,
structural/removal resistance
\\

C4 &
\cmark & \cmark & \xmark &
Human--device binding &
Joint key reconstruction, non-transferability,
clone and impersonation resistance
\\

C5 &
\xmark & \cmark & \cmark &
Device-bound functionality &
PUF compromise-to-unlock probability,
key-path security, physical-device binding
\\

C6 &
\cmark & \xmark & \cmark &
User-bound functionality &
Biometric-key reconstruction, impersonation resistance,
wrong-user activation
\\

C7 &
\cmark & \cmark & \cmark &
Human--device--function trust &
Legitimate activation, unauthorized activation,
cross-layer attack propagation
\\

\bottomrule
\end{tabularx}
\end{table*}
\section{Challenges, Limitations, and Open Research}
\label{sec:future}
\subsection{Compositional Security}
A secure biometric, secure PUF, and secure locking scheme do not automatically imply a secure composition. Interfaces can create new attacks: biometric errors alter PUF challenges; a PUF model can become a logic-key oracle; and a joint key can leak during configuration. Formal composition models and attack-chain benchmarks are largely missing.

\subsection{Revocable Human--Device Credentials}
Ownership changes require $(U_1,D)\rightarrow(U_2,D)$, and device replacement requires $(U,D_1)\rightarrow(U,D_2)$. Neither operation should expose reusable biometric information or require replacing the user's physiological trait. This requirement is consistent with the renewability/revocability objectives of biometric information protection in ISO/IEC~24745:2022. Domain-separated cancelable transforms, application-specific nonces, and reconfigurable hardware provide possible building blocks, but a general solution remains open.

\subsection{Joint Reliability Optimization}
Biometric and PUF stability are usually optimized independently. A joint design can instead select features/bits $S$ by
\begin{equation}
\label{eq:jointopt}
S^*=\arg\max_S\left[H(S)-\lambda P_e(S)-\nu L_{SCA}(S)\right],
\end{equation}
where $H$ is entropy, $P_e$ is reconstruction error, and $L_{SCA}$ is a leakage penalty. This formulation emphasizes the security--reliability trade-off without assuming that the two layers can be optimized independently.

\subsection{Learning-Resistant PUFs and Adaptive Attackers}
Strong-PUF evaluation should not stop at logistic regression or a single neural network. Modern work continues to improve both modeling attacks and modeling-resistant constructions \cite{alahmadi2024weakpuf,hongming2024attack}. Evaluation should vary attacker knowledge, CRP budget, architecture knowledge, environmental side information, and adaptive querying.

\subsection{Leakage-Aware End-to-End Hardware}
Constant-time ECG processing addresses one leakage dimension \cite{cordeiro2020timing}, but a C7 system also needs low-leakage PUF evaluation, KDF execution, key transport, and hardware configuration. Power/EM leakage, scan/test interfaces, DMA, caches, configuration buses, and debug ports must be included in the trust boundary.

\subsection{Secure Enrollment and PUF Modeling Assumptions}
Enrollment is the moment at which the biometric identity, device identity, and functional configuration become associated. A malicious enrollment can defeat all later authentication. Architectures that build or store a mathematical PUF model must explicitly protect that model and explain why model generation by the legitimate party does not create an equally effective adversarial path.

\subsection{Runtime Reconfiguration}
Static activation keys create long-lived attack targets. Session-specific keys $K_t=\KDF(B_u\parallel R_d\parallel N_t\parallel ID_{app})$ can support reconfigurable or partially reconfigurable hardware. The cost is reconfiguration latency, bitstream management, and more complex recovery logic.

\subsection{Standardized Benchmarks}
A useful benchmark should report, for the same end-to-end implementation, biometric EER/PAD, bio-key reliability/entropy, PUF reliability/uniqueness/modeling accuracy, logic-locking exact and approximate attack results, PPA, SCA/FI results, enrollment cost, and ownership-transfer/revocation behavior. Existing standards provide partial anchors---ISO/IEC~30136 for biometric template-protection evaluation and ISO/IEC~20897-2 for PUF testing---but they do not define a unified methodology for a composed biometric--PUF--obfuscation trust chain. Without such a benchmark, papers optimize different slices of the problem and are difficult to compare.


\FloatBarrier

\section{Conclusion}
\label{sec:conclusion}
Biometrics, PUFs, and hardware obfuscation establish different forms of identity: who the user is, which physical device is present, and which function is authorized to execute. A useful survey of their intersection must therefore go beyond listing techniques. It must expose the complete processing chain, the metrics used by each community, the cost of stabilization and protection, the assumptions behind the threat model, and the ways in which failures propagate across layers.

Biometric systems provide convenient human identity but must address noise, spoofing, privacy, and revocation. PUFs offer device-specific physical entropy but require environmental robustness, helper-data protection, and resistance to modeling and physical attacks. Logic locking provides functional authorization but faces oracle-guided, structural, approximate, and physical attacks while incurring PPA overhead. Their composition offers a path toward a human--silicon root of trust, but only if end-to-end reliability, key lifecycle, enrollment, side-channel leakage, and recovery are treated as first-class design constraints.

{\small
\bibliographystyle{IEEEtran}
\bibliography{references}
}

\end{document}